\documentclass[aps,pra,twocolumn,superscriptaddress]{revtex4-1}
\usepackage{tabularx}
\usepackage{dcolumn}
\usepackage{graphics}
\usepackage{bm}
\usepackage[dvips]{graphicx}
\usepackage[dvips]{rotating}
\usepackage[latin1]{inputenc}
\usepackage{dcolumn}
\usepackage{tabularx}
\usepackage{amsmath}
\usepackage{amssymb}
\usepackage{float}
\usepackage{natbib}
\usepackage[dvipsnames]{xcolor}
\begin{document}
\draft

\title{Magnetic Feshbach resonances in ultracold collisions between Cs and Yb atoms}

\author{B. C. Yang}
\affiliation{Joint Quantum Centre (JQC) Durham-Newcastle, Department of
Chemistry, Durham University, South Road, Durham, DH1 3LE, United Kingdom.}
\author{Matthew D. Frye}
\affiliation{Joint Quantum Centre (JQC) Durham-Newcastle, Department of
Chemistry, Durham University, South Road, Durham, DH1 3LE, United Kingdom.}
\author{A. Guttridge}
\affiliation{Joint Quantum Centre (JQC) Durham-Newcastle, Department of
Physics, Durham University, South Road, Durham, DH1 3LE, United Kingdom.}
\author{Jesus Aldegunde}
\affiliation{Departamento de Quimica Fisica, Universidad de Salamanca, 37008
Salamanca, Spain}
\author{Piotr S. \.{Z}uchowski}
\affiliation{Institute of Physics, Faculty of Physics, Astronomy and
Informatics, Nicolaus Copernicus University, ul.\ Grudziadzka 5/7, 87-100
Torun, Poland}
\author{Simon L. Cornish}
\email{s.l.cornish@durham.ac.uk} \affiliation{Joint Quantum Centre (JQC)
Durham-Newcastle, Department of Physics, Durham University, South Road, Durham,
DH1 3LE, United Kingdom.}
\author{Jeremy M. Hutson}
\email{j.m.hutson@durham.ac.uk} \affiliation{Joint Quantum Centre (JQC)
Durham-Newcastle, Department of Chemistry, Durham University, South Road,
Durham, DH1 3LE, United Kingdom.}

\date{\today}

\begin{abstract}
We investigate magnetically tunable Feshbach resonances in ultracold collisions
between ground-state Yb and Cs atoms, using coupled-channel calculations based
on an interaction potential recently determined from photoassociation
spectroscopy. We predict resonance positions and widths for all stable isotopes
of Yb, together with resonance decay parameters where appropriate. The
resonance patterns are richer and more complicated for fermionic Yb than for
spin-zero isotopes, because there are additional level splittings and couplings
due to scalar and tensorial Yb hyperfine interactions. We examine collisions
involving Cs atoms in a variety of hyperfine states, and identify resonances
that appear most promising for experimental observation and for
magnetoassociation to form ultracold CsYb molecules.
\end{abstract}

\maketitle

\section{Introduction}

Magnetic Feshbach resonances are a valuable tool for tuning the scattering
length by varying an external magnetic field, and have found a wide range of
applications in studying and controlling ultracold gases \cite{RMP_Chengchin}.
Great progress has been achieved in the exploration of Feshbach resonances for
pairs of alkali-metal atoms. One significant achievement is the formation of
ultracold molecules by adiabatically ramping the magnetic field across a
zero-energy Feshbach resonance \cite{Li2_Hulet_2003PRL, Li2_pure_2003PRL,
Li2_Salomon_2003PRL, Na2_2003PRL, K2_2003Nature, Rb2_2002Nature, Rb2_2004PRL,
Cs2_pure_2003Science, NaLi_2012PRA}, known as magnetoassociation
\cite{RMP_Magnetoassociation, Hutson_Soldan}. The resulting weakly bound dimers
can be transferred to their absolute ground states by stimulated Raman
adiabatic passage \cite{Rb2_STIRAP_OL_2008PRL, KRb_STIRAP_2008Science,
Cs2_STIRAP_OL_2010NP, RbCs_STIRAP_Innsbruck, RbCs_STIRAP_Durham,
NaK_STIRAP_MIT, NaRb_STIRAP_CUHK, NaLi_STIRAP_2017PRL}. These ultracold
molecules promise diverse applications in fields from ultracold chemistry to
precision measurement, due to their rich internal degrees of freedom and
complex interactions compared to ultracold atoms \cite{Lincoln_Carr_review,
DDI_ChemRev}. In particular, the inherent electric dipole moment of ultracold
polar molecules makes them valuable in studying quantum dipolar matter
\cite{DDI_Boson_Pfau, DDI_molecules_NP} and for applications in quantum
computation and simulation \cite{QComp_KCs_PRL, RbCs_CaF_QSimul}.

There is currently great interest in ultracold mixtures of alkali-metal and
closed-shell ($^1$S) atoms \cite{YbRb_01photon_2009PRA, YbRb_02photon_2011PCCP,
LiYb_mixture_Takahashi_2011PRL, LiYb_mixture_Gupta_2011PRL,
LiYb_mixture_Gupta_2011PRA, Alk1S_2010PRL_RbSr, Alk1S_2012PRL_LiYb,
Alk1S_2013PRA_RbYb_CsYb, RbSr_mixture_2013PRA, CsYb_MOT, CsYb_ZMSL,
CsYb_Thermalization}. The molecules formed from these atoms have $^2\Sigma$
ground states with unpaired electron spin. They therefore have both electric
and magnetic dipole moments, and provide a new platform for studying lattice
spin models in many-body physics \cite{QComp_LatSpin_NP}. They may also be
valuable in searches for the electric dipole moment of the electron
\cite{Electron_EDM_CsYb}. However, magnetoassociation in such mixtures will be
challenging because the Feshbach resonances are expected to be narrow. This is
because the lack of structure of a $^1$S atom removes the strong couplings that
cause many wide resonances in alkali+alkali systems; the strongest source of
coupling in alkali+$^1$S systems is the weak dependence of hyperfine coupling
on interatomic distance \cite{Alk1S_2010PRL_RbSr, Alk1S_2012PRL_LiYb}.
Nevertheless, Feshbach resonances have recently been observed in an ultracold
Rb+Sr mixture \cite{Alk1S_2018NP_RbSr}. There is now great hope that it will be
possible to form ultracold open-shell molecules by magnetoassociation at these
resonances.

We have studied ultracold Cs+Yb mixtures both experimentally and theoretically
\cite{Alk1S_2013PRA_RbYb_CsYb, CsYb_MOT, CsYb_ZMSL, CsYb_Thermalization,
CsYb_01photon}. The merits of this system include the existence of seven stable
isotopes of Yb, including five spin-zero bosons and two fermions. Because of
the large mass of Cs, significant variation of the atom-pair reduced mass can
be achieved by choosing different isotopes of Yb. This produces a wide variety
of Feshbach resonances with substantially different properties for different
isotopes \cite{Alk1S_2013PRA_RbYb_CsYb}. However, the predictions of Ref.\
\cite{Alk1S_2013PRA_RbYb_CsYb} were limited because, at that time, the
ground-state interaction potential was not known accurately enough to predict
scattering lengths for specific isotopic combinations.

\begin{figure}[tbp]
\includegraphics[width=\columnwidth]{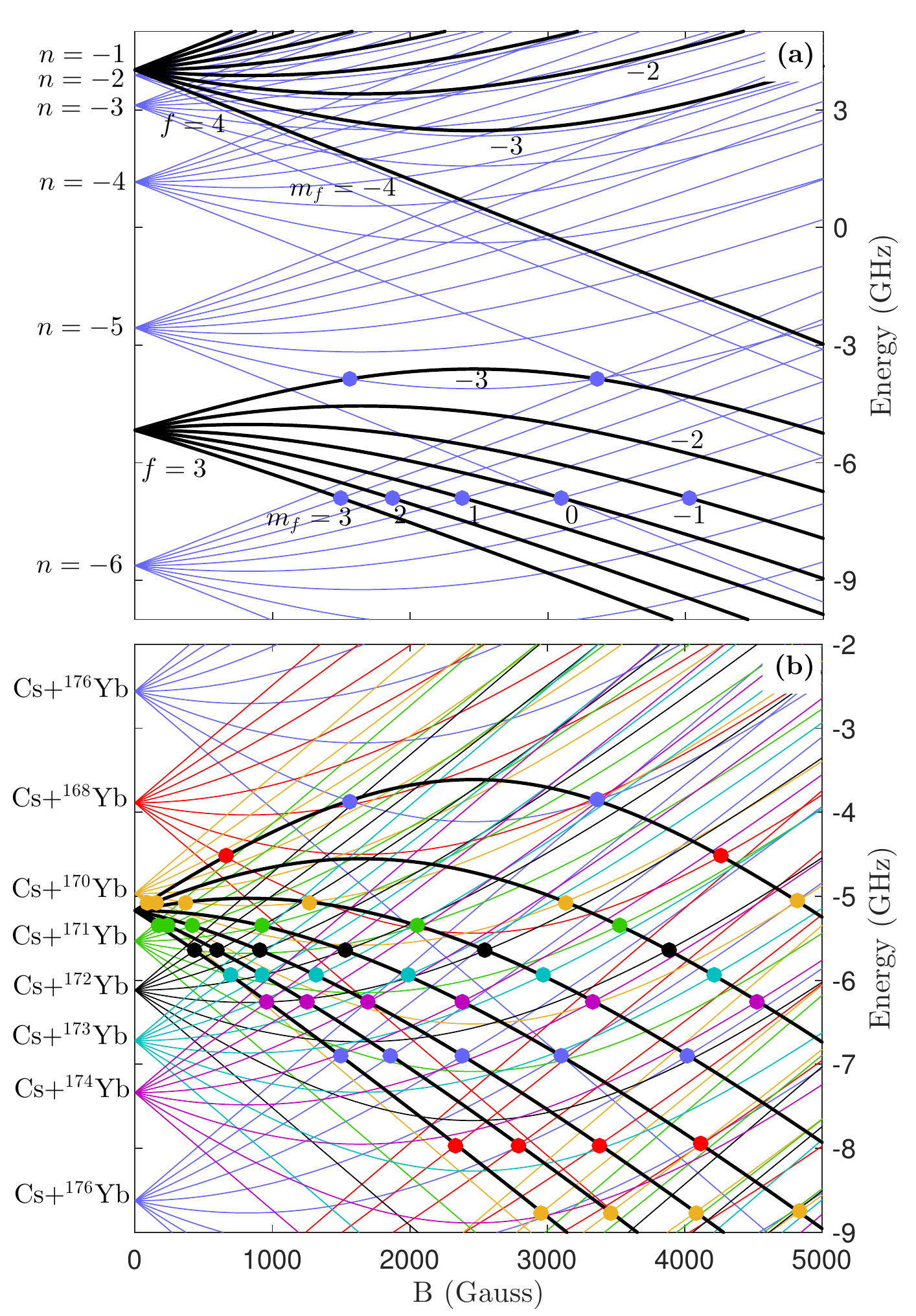}
\caption{\label{fig:level_intro}
(Color online) Near-threshold bound states (thin colored lines) crossing atomic
thresholds (thick black lines) as a function of magnetic field. The solid
circles mark the Feshbach resonances caused by the dependence of the Cs
hyperfine coupling on the internuclear distance. (a) The hyperfine + Zeeman
splittings of the atomic threshold and molecular bound states for the example
of Cs+$^{176}$Yb. The atomic levels are labeled by quantum numbers $f$ and
$m_f$ as discussed in the text. Only molecular levels from the upper hyperfine
manifold ($f=4$) are shown, labeled by the vibrational number $n$. (b)
molecular levels for $n= -5$ and $-6$ for all isotopic combinations, with
Feshbach resonance positions at crossings with atomic states in the lower
hyperfine manifold. }
\end{figure}

In recent work, we have measured the binding energies of near-threshold bound
states for several isotopologs of CsYb and determined the ground-state
electronic potential \cite{CsYb_02photon}. This allows us to make specific
predictions for the positions and widths of Feshbach resonances. Figure
\ref{fig:level_intro} shows the atomic thresholds for Cs as a function of
magnetic field, and the near-threshold energy levels of CsYb predicted for the
ground-state potential of Ref.\ \cite{CsYb_02photon}. Because there is only a
single electronic state and the hyperfine coupling is weakly dependent on
distance, the molecular levels are essentially parallel to the threshold that
supports them \cite{Alk1S_2010PRL_RbSr, Alk1S_2013PRA_RbYb_CsYb}. Feshbach
resonances due to the Cs hyperfine coupling are predicted at the crossings
indicated by colored dots. The figure also shows how different choices of Yb
isotope shift the near-threshold states and strongly affect the resonance
positions.

In this paper we perform coupled-channel calculations to identify, locate, and
characterize Feshbach resonances in ultracold collisions between Cs and Yb
atoms. Our main focus is to understand the physics behind the properties of
Feshbach resonances in this system and to establish which Feshbach resonances
are promising for experimental observation and molecule formation. In Sec.\
\ref{sec:theory} we introduce the underlying theory of these Feshbach
resonances: the coupling terms in the Hamiltonian which cause them, the methods
we use to characterize them, and the framework we use to understand the
results. In Sec.\ \ref{sec:mechanisms} we use Cs+$^{173}$Yb as an example
system to discuss the effects of the different coupling mechanisms and the
general characteristics of the different resonances they cause. In Sec.\
\ref{sec:predictions} we identify promising resonances for observation and
magnetoassociation for various isotopic combinations of Cs+Yb, taking account
of experimental considerations. Comprehensive results for resonances of all
isotopic combinations are provided in Supplemental Material.

\section{Theoretical background} \label{sec:theory}

We consider the ultracold scattering between $^{133}$Cs($^2$S) and Yb($^1$S)
atoms. The Hamiltonian $\hat{H}$ can be written \cite{Alk1S_2012PRL_LiYb}
\begin{equation}
\label{full_H}
\hat{H} =\frac{\hbar^2}{2\mu}\left[-\frac{1}{R}\frac{d^2}{dR^2}R
+\frac{\hat{L}^2}{R^2}\right]+\hat{H}_\textrm{Cs}+\hat{H}_\textrm{Yb}+\hat{U}(R),
\end{equation}
where $R$ is the internuclear distance, $\mu$ is the reduced mass, and $\hbar$
is the reduced Planck constant. $\hat{L}$ is the two-atom rotational angular
momentum operator, with quantum number $L$ and projection $M_L$.
$\hat{H}_\textrm{Cs}$ and $\hat{H}_\textrm{Yb}$ are the Hamiltonians for the
separated single atoms, which are independent of $R$ and contain hyperfine
coupling and Zeeman terms,
\begin{align}
\hat{H}_\textrm{Cs} &= \zeta_\textrm{Cs} \hat{i}_\textrm{Cs} \cdot
\hat{s}+(g_\textrm{Cs} \hat{i}_{\textrm{Cs},z}+g_\textit{s} \hat{s}_{z}) \mu_{\rm B} B,\\
\label{Yb_Zeeman}
\hat{H}_\textrm{Yb} &= g_\textrm{Yb} \hat{i}_{\textrm{Yb},z} \mu_{\rm B} B.
\end{align}
Here, $B$ is a magnetic field oriented along the $z$ axis, and $\mu_{\rm B}$ is
the Bohr magneton. $\zeta_\textrm{Cs}$ is the hyperfine coupling constant for
the Cs atom. $\hat{i}_\textrm{Cs}$, $\hat{i}_\textrm{Yb}$ and $\hat{s}$ are the
nuclear and electron spin operators, with projections on the $z$ axis
$\hat{i}_{\textrm{Cs},z}$, $\hat{i}_{\textrm{Yb},z}$, and $\hat{s}_{z}$; their
corresponding $g$ factors are $g_\textrm{Cs}$, $g_\textrm{Yb}$, and $g_s$,
respectively. The specific values of $\zeta_\textrm{Cs}$,  $g_\textrm{Cs}$, and
$g_s$ in this work are taken from Ref.\ \cite{alkali_hyperfine_1977RMP} and
those for $g_\textrm{Yb}$ are obtained from the shielded magnetic moments
(without diamagnetic correction) of Ref.\ \cite{Yb_Nuclear_Moment}.

The interaction operator $\hat{U}(R)$ is divided into the electronic
interaction potential $V_\textrm{elec}(R)$ and spin-dependent terms
$\hat{V}_\textrm{spin}(R)$. The electronic potential is by far the strongest
interaction, and almost completely determines the bound states and non-resonant
scattering in each channel. However, it cannot change the electron or nuclear
spins and so does not couple different channels or cause Feshbach resonances.
We use the ground-state interaction potential fitted to two-photon spectroscopy
in Ref.\ \cite {CsYb_02photon}, which provides an accurate representation of
the near-threshold bound states that produce Feshbach resonances. The
scattering length and the energy of the highest bound state are given in Table
\ref{table:scatln} for Cs interacting with each isotope of Yb on this potential
(without any internal structure on either collision partner).

\begin{table}[tbp]
  \centering
\caption{\label{table:scatln} Scattering length and energy of the highest bound
state for Cs interacting with each isotope of Yb on the potential
$V_\textrm{elec}(R)$ of Ref.\ \cite{CsYb_02photon}.}
\begin{ruledtabular}
    \begin{tabular}{ccc}
   Mixture			& $a\ (a_0)$	& $E_\textrm{b}(n=-1)/h$ (MHz)	\\
\hline
Cs+$^{168}$Yb		& 165.98		& 3.70	\\
Cs+$^{170}$Yb		& 96.24		& 15.6	\\
Cs+$^{171}$Yb		& 69.99		& 25.8	\\
Cs+$^{172}$Yb		& 41.03		& 39.5	\\
Cs+$^{173}$Yb		& 1.0			& 57.0	\\
Cs+$^{174}$Yb		& $-74.8$		& 78.7	\\
Cs+$^{176}$Yb		& 798		& 0.0513	\\
\end{tabular}
\end{ruledtabular}
\end{table}

\subsection{Spin-dependent terms} \label{sec:coupling}

The systems considered here lack the strong couplings that cause wide Feshbach
resonances for pairs of alkali-metal atoms, due to differences between singlet
and triplet potentials and electron spin-spin couplings. Instead, couplings
between different channels are caused by the change in hyperfine interactions
due to the proximity of the two atoms. The operator $\hat{V}_\textrm{spin}(R)$
may be written \cite{Jesus_Hyperfine},
\begin{align}
\hat{V}_\textrm{spin}(R) &= \Delta\zeta_\textrm{Cs}(R) \hat{i}_\textrm{Cs} \cdot \hat{s}
+ \Delta\zeta_\textrm{Yb}(R) \hat{i}_\textrm{Yb} \cdot \hat{s}  \nonumber \\
& \quad +t_\textrm{Yb}(R) \sqrt{6}~T^2(\hat{i}_\textrm{Yb}, \hat{s})\cdot T^2(C) \nonumber\\
& \quad +t_\textrm{Cs}(R) \sqrt{6}~T^2(\hat{i}_\textrm{Cs}, \hat{s})\cdot T^2(C) \nonumber\\
& \quad +e{\bm Q}_\textrm{Yb}\cdot{\bm q}_\textrm{Yb}(R) \nonumber\\
& \quad +e{\bm Q}_\textrm{Cs}\cdot{\bm q}_\textrm{Cs}(R) + \gamma(R) \hat{s} \cdot \hat{L}.
\label{Vcpl_R}
\end{align}
The first two terms represent the scalar contact interaction between the
electron and nuclear spins, while the third and fourth terms represent the
corresponding dipolar interaction. Here $T^2$ indicates a spherical tensor of
rank 2; $T^2(C)$ has components $C^2_q(\theta,\phi)$, where $C$ is a
renormalised spherical harmonic and $\theta,\phi$ are the polar coordinates of
the internuclear vector. The fifth and sixth terms represent the interaction
between the nuclear electric quadrupole tensor $e\bm{Q}_j$ of nucleus $j$ and
the distance-dependent electric field gradient tensor ${\bm q}_j(R)$ at the
nucleus, due to the electrons. The final term represents the interaction
between the electron spin and the molecular rotation.

The first three terms in Eq.\ \ref{Vcpl_R} are the ones that are principally
responsible for Feshbach resonances in CsYb and similar systems. We refer to
them as mechanisms I, II, and III, respectively; each can be written as the
product of a purely $R$-dependent term $\omega_x(R)$ and a purely
spin-dependent term $\hat{\Omega}_x$ that is different for each of $x={}$I, II,
and III.

Mechanism I is due to the variation in hyperfine coupling on the Cs atom,
$\hat{\Omega}_\textrm{I}=\hat{i}_\textrm{Cs}\cdot \hat{s}$. This arises because
the approaching Yb atom pulls electron-spin density away from the Cs nucleus,
thereby reducing the strength of the hyperfine interaction. This coupling
mechanism was first proposed by \.Zuchowski, Aldegunde, and Hutson
\cite{Alk1S_2010PRL_RbSr} for Rb+Sr and was investigated extensively by Brue
and Hutson for alkali-metal + Yb systems \cite{Alk1S_2013PRA_RbYb_CsYb}. As it
relies only on the Cs nuclear spin, it exists for all isotopic combinations of
Cs+Yb.

Mechanism II is due to the variation in hyperfine coupling on the Yb atom,
$\hat{\Omega}_\textrm{II}=\hat{i}_\textrm{Yb}\cdot \hat{s}$. This mechanism is
complementary to mechanism I: as electron-spin density is pulled away from the
Cs nucleus, some of it comes into contact with the Yb nucleus, where it can
interact with a nuclear spin. This mechanism was first proposed by Brue and
Hutson \cite{Alk1S_2012PRL_LiYb}. It exists only for Yb isotopes with a
non-zero nuclear spin, so only for $^{171}$Yb and $^{173}$Yb.

Mechanism III is due to the tensor, or anisotropic, hyperfine coupling on the
Yb atom, $\hat{\Omega}_\textrm{III}=\sqrt{6}T^2(\hat{i}_\textrm{Yb},
\hat{s})\cdot T^2(C)$. The approach of the Cs atom breaks the spherical
symmetry of the electron density around the Yb nucleus and allows a dipolar
coupling that can cause resonances due to $L=2$ bound states in s-wave
scattering. This mechanism was briefly considered by Brue and Hutson
\cite{Alk1S_2012PRL_LiYb} but they ultimately neglected it; nevertheless,
resonances caused by this mechanism were later observed in Rb+$^{87}$Sr
\cite{Alk1S_2018NP_RbSr}. Like mechanism II, this mechanism relies on the Yb
nuclear spin so exists only for $^{171}$Yb and $^{173}$Yb.

The fourth term in Eq.\ \ref{Vcpl_R}, involving $t_\textrm{Cs}$, is analogous
to the third and may formally be considered as contributing to mechanism III.
However, it is very weak in CsYb, as discussed below. The quadrupole term
involving $Q_\textrm{Yb}$ does not generally produce resonances, but may cause
significant level shifts for levels of Cs$^{171}$Yb and Cs$^{173}$Yb with
$L>0$, as described in section \ref{sec:Mech_III}. The quadrupole term
involving $Q_\textrm{Cs}$ can in principle cause resonances due to $L=2$ bound
states, but is very weak in CsYb. The spin-rotation term $\gamma(R) \hat{s}
\cdot \hat{L}$ has no matrix elements involving $L=0$ states so does not cause
resonances in s-wave scattering. All terms except that involving
$t_\textrm{Cs}$ are included where applicable in the coupled-channel
calculations described below.

\subsection{Electronic structure calculations of spin-dependent coefficients}

We have calculated values of the scalar hyperfine coupling coefficients
$\zeta_{\rm Cs}(R)$ and $\zeta_{\rm Yb}(R)$, the corresponding tensor
coefficients $t_{\rm Cs}(R)$ and $t_{\rm Yb}(R)$, and the nuclear quadrupole
coupling coefficients $(eQq)_{\rm Cs}(R)$ and $(eQq)_{\rm Yb}(R)$. We have also
calculated the electron $g$-tensor anisotropy $\Delta g_\bot(R)$, which is
related to the spin-rotation coefficient $\gamma(R)$ \cite{Jesus_Hyperfine}. We
carried out density-functional (DFT) calculations using the Amsterdam Density
Functional (ADF) package \cite{ADF1,ADF3} as described in Ref.\
\cite{Jesus_Hyperfine}, at 40 distances from $R=3.8$~\AA\ to 20~\AA. The
coefficients for $^{171}$Yb are obtained from those for $^{173}$Yb by scaling
using nuclear $g$-factors, nuclear quadrupole moments and molecular rotational
constants as described in Ref.\ \cite{Aldegunde:polar:2008}.

Aldegunde and Hutson \cite{Jesus_Hyperfine} concluded that the B3LYP functional
\cite{Beck93, Stephens:1994} gives good accuracy for hyperfine coupling
coefficients in $^2\Sigma$ molecules, and that spin-unrestricted calculations
are slightly more accurate than restricted calculations when the two results
are similar. We obtained similar results from restricted and unrestricted
calculations, so we report the unrestricted results here. The one exception to
this is the coefficient $t_\textrm{Cs}(R)$, which is so small that the
differences between the restricted and unrestricted results are comparable to
their absolute magnitude. We consider these results to be consistent with zero,
so do not report $t_\textrm{Cs}(R)$ and exclude the corresponding term from our
coupled-channel calculations.

For all the coefficients, the values from DFT calculations behave irregularly
inside the zero-energy inner turning point $\sigma_0$, which is at 4.1~\AA\ for
CsYb. The irregularities probably occur because different electronic states mix
strongly in the region of the repulsive wall. Since the resonance properties of
interest here are insensitive to the behavior of the couplings inside the inner
turning point, we have fitted functional forms to the points at $R\ge 4.0$~\AA.

The scalar hyperfine coupling coefficients $\Delta\zeta_{\rm Cs}(R)=\zeta_{\rm
Cs}(R)-\zeta_{\rm Cs}$ and $\Delta\zeta_{\rm Yb}(R)=\zeta_{\rm Yb}(R)$ are both
negative for all $R>\sigma_0$, but both of them show positive curvature
slightly outside $\sigma_0$. The same is true for the quadrupole coupling
coefficients $(eQq)_{\rm Cs}(R)$ and $(eQq)_{\rm Yb}(R)$. For consistency with
Brue and Hutson \cite{Alk1S_2013PRA_RbYb_CsYb}, we have chosen to represent
these coefficients with Gaussian functions, $A_0\exp[-a(R-R_c)^2]$. However,
there is no sign of such curvature for $t_\textrm{Yb}$ or $\gamma(R)$, and for
these we have used simple decaying exponentials $A_0\exp(-b(R-\sigma_0))$, with
$\sigma_0$ fixed at 4.1~\AA. The resulting parameters are given in Table
\ref{parameters}.

The calculated function $\Delta\zeta_{\rm Cs}(R)$ predicts that the $f=4$,
$n=-7$ level of Cs$^{174}$Yb is bound by 11 MHz more than the corresponding
$f=3$ level. This may be compared with an experimental shift of $10\pm3$~MHz
from 2-photon photoassociation spectroscopy \cite{CsYb_02photon,
Guttridge-unpub:2018}.

\begin{table}[tbp]
  \centering
\caption{Parameters for the $R$-dependence of the spin-dependent
coefficients.}
\label{parameters}
\begin{ruledtabular}
\begin{tabular}{rrrr}
  &   $A_0$ (MHz)   &  $a$ (\AA$^{-2})$  &  $R_{\rm c}$ (\AA) \\  \cline{2-4}
  $\Delta\zeta$ ($^{133}$Cs) & $-241$ & $0.154$ & $3.33$ \\
  $\Delta\zeta$ ($^{173}$Yb) & $-126$ & $0.144$ & $3.42$ \\
  $\Delta\zeta$ ($^{171}$Yb) & $457$ & $0.144$ & $3.42$ \\
  $eQq$ ($^{133}$Cs) & $0.227$ & $0.256$ & $3.28$ \\
  $eQq$ ($^{173}$Yb) & $-601$ & $0.249$ & $3.32$ \\
  \hline
  & $A_0$ (MHz) & $b$ (\AA$^{-1}$) & $\sigma_0$ (\AA) \\ \cline{2-4}
  $t$ ($^{173}$Yb) & $-24.5$ & $0.953$ & 4.1 \\
  $t$ ($^{171}$Yb) & $88.9$ & $0.953$ & 4.1 \\
  $\gamma$  & $21.7$ & $1.58$ & 4.1 \\
\end{tabular}
\end{ruledtabular}
\end{table}

\subsection{Magnetic Feshbach resonances}
\label{scattering_length}

A magnetic Feshbach resonance occurs when a molecular bound state is tuned
across an atomic scattering threshold by varying an applied magnetic field. For
an isolated resonance without inelastic decay, the scattering length $a(B)$ has
a characteristic pole at the resonance position $B_\textrm{res}$
\cite{Moerdijk:1995},
\begin{equation}
\label{eq:a_res}
a(B)=a_\textrm{bg}\left(1-\frac{\Delta}{B-B_\textrm{res}}\right),
\end{equation}
where $a_\textrm{bg}$ is the background scattering length. The resonance width
$\Delta$ can be artificially large if $a_\textrm{bg}$ is particularly small. A
better measure of the strength of the resonant pole is the product
$a_\textrm{bg}\Delta$, which provides a measure of the observability of the
resonance in 3-body loss spectroscopy and is proportional to the rate of the
field sweep needed to achieve adiabatic passage in magnetoassociation
\cite{Mies:Feshbach:2000, Julienne:2004}. However,
$a_\textrm{bg}\Delta$ has inconvenient dimensions for qualitative discussion;
in order to maintain a measure with dimensions of magnetic field, we define a
normalized width
\begin{equation}
\bar{\Delta}=a_\textrm{bg}\Delta/\bar{a},
\label{eq:delta-bar}
\end{equation}
where $\bar{a}=(2\mu C_6/\hbar^2)^{1/4}\times 0.4779888\dots$ is the mean
scattering length of Gribakin and Flambaum \cite{Gribakin:1993} and ranges from
83.62 $a_0$ for Cs+$^{168}$Yb to 84.05 $a_0$ for Cs+$^{176}$Yb.

When inelastic decay occurs, the scattering length becomes complex, with the
imaginary part describing inelastic loss \cite{Balakrishnan:scat-len:1997}.
Near a resonance, the scattering length no longer has a pole, but instead both
real and imaginary parts show an oscillation; this may be written
\cite{Hutson:res:2007}
\begin{equation}
\label{eq:a_decay}
a(B)=a_\textrm{bg}+\frac{a_\textrm{res}}{2(B-B_\textrm{res})/ \Gamma^\textrm{inel}_B+i},
\end{equation}
where $a_\textrm{res}$ is a resonant scattering length that characterizes the
oscillation. In general, both $a_\textrm{bg}$ and $a_\textrm{res}$ are complex,
but the weak background inelasticity for CsYb means that they are nearly real
and we will neglect their complex parts. If $|a_\textrm{res}|$ is large then
the oscillation in the real part of the scattering length is large and
pole-like, similar to the case without decay. $\Gamma^\textrm{inel}_B$ is a
decay width in field; the decay width in energy is given by
$\Gamma^\textrm{inel}_E=\Gamma^\textrm{inel}_B\delta\mu$, where $\delta\mu$ is
the magnetic moment of the bare resonant state relative to the atomic state.
When inelasticity is present, the molecule formed by magnetoassociation can
decay (predissociate) with lifetime $\tau=\hbar/\Gamma^\textrm{inel}_E$. We
define the width $\Delta$ for a decayed resonance through
\begin{equation}
\label{eq:width_inelastic}
a_\textrm{bg}\Delta=-a_\textrm{res}\Gamma^\textrm{inel}_B/2.
\end{equation}
This gives the same behavior in the wings as for an undecayed resonance of the
same width \cite{Matthew_Converge}.

\subsection{Coupled-channel calculations}

To locate and characterize the Feshbach resonances, we use coupled-channel
bound-state and scattering calculations. The wavefunction is expanded in an
uncoupled basis set
\begin{equation}
|s, m_s\rangle |i_\textrm{Cs}, m_{i,\textrm{Cs}}\rangle |i_\textrm{Yb},
m_{i,\textrm{Yb}}\rangle |L, M_L\rangle.
\label{eq:basis}
\end{equation}
Here $s$, $i_\textrm{Cs}$ and $i_\textrm{Yb}$ are quantum numbers for the
electron and nuclear spin angular momenta and $m_s$, $m_{i,\textrm{Cs}}$ and
$m_{i,\textrm{Yb}}$ are the corresponding projections onto the axis of the
magnetic field. The only conserved quantum numbers are the total angular
momentum projection $M_\textrm{tot}=m_s+m_{i,\textrm{Cs}}+m_{i,\textrm{Yb}}+M_L$ and the total
parity $(-1)^L$. Basis sets are constructed including all functions of the
required $M_\textrm{tot}$ and parity +1, including functions up to $L_{\rm
max}$. Different situations require $L_{\rm max}=0$, 2 or 4, as described
below.

The 5 bosonic isotopes of Yb all have zero nuclear spin, $i_\textrm{Yb}=0$, and
the two fermions, $^{171}$Yb and $^{173}$Yb, have $i_\textrm{171Yb}=1/2$ and
$i_\textrm{173Yb}=5/2$, respectively. For the Cs atom, $i_\textrm{Cs}=7/2$ and
$s=1/2$; these can be coupled to give a resultant $f=3$ or 4. The corresponding
projection $m_f$ is conserved by $\hat{H}_\textrm{Cs}$, but $f$ is not except
at $B=0$. Nonetheless, we label states by the value of $f$ that they correlate
with at $B=0$ \footnote{In Cs, the large hyperfine splitting means that $f$
remains a relatively good quantum number up to quite high fields, including
most fields of experimental interest. However, lighter alkali-metal atoms have
smaller hyperfine splittings so a different choices of labels may be
appropriate.}.

The resulting coupled equations are constructed and solved for bound states
using the \textsc{bound} and \textsc{field} programs \cite{bound+field:2019,
mbf-github:2019} and for scattering using the \textsc{molscat} program
\cite{molscat:2019, mbf-github:2019}. In the short-range region, $3.5\textrm{
\AA}<R<25$~\AA, solutions are propagated using the diabatic log-derivative
method of Manolopoulos \cite{LDMD1986, LDMD1993} with a fixed step size
0.001~\AA; in the long-range region, $25\ \textrm{\AA} \leq R\leq
R_\textrm{max}$, the log-derivative Airy propagator of Alexander and
Manolopoulos is applied with a variable step size \cite{Alexander1987}. This
allows efficient propagation to very large values of $R_\textrm{max}$. The
calculations are converged with respect to integration range and step size.

For scattering calculations, log-derivative solutions are propagated outwards
from short range to a distance $R_{\rm max}=50,000$~\AA\ at long range. Since
the basis functions (\ref{eq:basis}) are not eigenfunctions of the
separated-atom Hamiltonian, the resulting log-derivative matrix at $R_{\rm
max}$ is transformed to the separated-atom basis set and then matched to
asymptotic boundary conditions to obtain the ${\bf K}$ matrix and then the
scattering {\bf S} matrix. The scattering length is obtained as $a(k) =
(ik)^{-1} (1-S_{00})/(1+S_{00})$, where $k=\sqrt{2\mu E}/\hbar$ is the
wavevector and $S_{00}$ is the diagonal S-matrix element in the incoming
channel. The kinetic energy in the incoming channel is set to be
$E=100$~nK$\times k_\textrm{B}$, where $k_\textrm{B}$ is the Boltzmann
constant; this energy is low enough that the resulting scattering length has
essentially reached its zero-energy value. We use the algorithms of Frye and
Hutson \cite{Matthew_Converge} to locate and characterize the Feshbach
resonances in the calculated scattering lengths.

For bound-state calculations, one log-derivative solution $\boldsymbol{Y}_{\rm
out}(R)$ is propagated outwards from short range, and another
$\boldsymbol{Y}_{\rm in}(R)$ is propagated inwards from $R_{\rm max}=200$~\AA,
until both reach a matching point $R_{\rm match}$ in the classically allowed
region. At a bound-state energy, the matching matrix $\boldsymbol{Y}_{\rm
out}(R_{\rm match})-\boldsymbol{Y}_{\rm in}(R_{\rm match})$ has zero
determinant and one of its eigenvalues is zero \cite{Jeremy_BDmatching1994}.
\textsc{bound} locates eigenenergies by varying the energy of the calculation
at fixed magnetic field until the matching condition is met. \textsc{field}
operates similarly, but varies the magnetic field at fixed energy relative to
threshold. This approach allows us to converge efficiently and accurately on
bound-state energies and on magnetic fields at which bound states cross
threshold.

\subsection{Fermi's golden rule}

Accurate Feshbach resonance widths can be obtained from coupled-channel
scattering calculations, but such calculations do not provide much insight. We
therefore use an analysis based on Fermi's golden rule to understand our
results. This gives an expression for the resonance width in terms of the
matrix element of the coupling operator $\hat{V}_\textrm{spin}(R)$ between the
single-channel scattering state $| \alpha k \rangle$, which is labeled by a
channel index $\alpha$ and wavevector $k$ and normalised to a $\delta$-function
of energy, and the bound state $|\alpha' n\rangle$, where $n$ is the
vibrational quantum number relative to threshold. Brue and Hutson showed that
the width can be written \cite{Alk1S_2013PRA_RbYb_CsYb}
\begin{equation}
\label{Fermi_rule}
\Delta=\frac{\pi}{ka_\textrm{bg} \delta\mu_\text{res}}
\langle n| \omega_x(R) |k\rangle_R^2 \langle\alpha'|\hat{\Omega}_x|\alpha\rangle_\textrm{spin}^2,
\end{equation}
where the matrix element has been separated into a radial component $\langle
\cdots \rangle_R$, and a spin component $\langle \cdots \rangle_\textrm{spin}$.

The separation of the two components of the matrix element allows a clear
interpretation of the factors that influence the resonance widths. The spin
component $\langle\alpha'|\hat{\Omega}_x|\alpha\rangle$, which was denoted
$I_{m_{f,a}}(B)$ for mechanism I in Ref.\ \cite{Alk1S_2013PRA_RbYb_CsYb},
describes how the coupling strength depends on the spin states that are coupled
and how it varies with magnetic field. The radial component $\langle n|
\omega_x(R) |k\rangle$ takes account of the binding energy of the bound state
and the background scattering length in the incoming channel. Near threshold,
$\langle n| \omega_x(R) |k\rangle$ is proportional to $k^{1/2}$, so that
$\bar{\Delta}$ is independent of energy to first order.

The golden rule approach can be used as an approximate method of calculating
widths, but in this paper we use it only as an interpretative tool. All widths
presented are from coupled-channel calculations.

\section{Coupling mechanisms} \label{sec:mechanisms}

In this section, we explore the resonances caused by the three principal
coupling mechanisms described in Sec.\ \ref{sec:coupling}. We focus on the
general patterns of the resonance positions and widths, rather than the
specific predictions, which are given in Sec.\ \ref{sec:predictions}. We also
consider inelastic decay.

We take Cs+$^{173}$Yb as our example system in this section, although the
analysis is relevant to other isotopologs and other systems formed from an
alkali-metal atom and a closed-shell atom, such as Rb+Sr. The scattering length
for $V_\textrm{elec}(R)$ is very small for Cs+$^{173}$Yb, so that
$a_\textrm{bg}$ can vary substantially between resonances, and it is important
to use the normalized width $\bar{\Delta}$ (Eq.\ \ref{eq:delta-bar}) rather
than $\Delta$ itself as the measure of resonance strength.

\subsection{Mechanism \textrm{I}}

Resonances caused by mechanism I have been investigated by Brue and Hutson
\cite{Alk1S_2013PRA_RbYb_CsYb}. However, at that time the binding energies and
scattering lengths for Cs+Yb were unknown, so they could study only the general
properties.

The operator $\hat{\Omega}_\textrm{I}=\hat{i}_\textrm{Cs}\cdot \hat{s}$
responsible for mechanism I produces couplings with selection rule $\Delta
m_f=0$, where the notation $\Delta x=x_\textrm{bound}-x_\textrm{scat}$
indicates the change in quantum number $x$ between the incoming scattering
state and the resonant bound state. Since there is only one atomic state of
each $m_f$ for each $f$, mechanism I couples molecular bound states to atomic
scattering states only if they have different values of $f$. Each bound state
is essentially parallel to the atomic threshold that supports it, and Fig.\
\ref{fig:level_intro}(b) shows the resulting crossing diagram. The molecular
states that produce Feshbach resonances by mechanism I correspond to $f=4$, so
at the energy of the $f=3$ thresholds they are bound by approximately the Cs
hyperfine splitting. The bound states are therefore sparsely distributed in
energy and the corresponding resonances are sparsely distributed in magnetic
field. The matrix element $\langle f m_f |\hat{\Omega}_\textrm{I}|f'
m_f\rangle$ goes linearly to zero as $B\rightarrow 0$, so the resulting
resonance widths $\bar{\Delta}$ are proportional to $B^2$ at low fields
\cite{Alk1S_2013PRA_RbYb_CsYb}. Resonances with usefully large widths thus
exist at accessible magnetic fields only if a bound state for $f=4$
accidentally falls close to the $f=3$ thresholds.

Resonances caused by mechanism I are present for both bosonic and fermionic
isotopes of Yb. For bosonic isotopes, Yb hyperfine couplings are absent. Since
there are no significant anisotropic couplings in this case, we use
calculations with $L_{\rm max}=0$. However, for fermionic isotopes ($^{171}$Yb
and $^{173}$Yb) with nonzero nuclear spin, the hyperfine coupling terms
corresponding to mechanisms \textrm{II} and \textrm{III} can alter the
resonance widths produced by mechanism I alone and in some cases introduce
inelastic decay. These effects are discussed in the following subsections.

\subsection{Mechanism \textrm{II}}
\label{Mech_II}

Mechanism \textrm{II} is due to the scalar hyperfine coupling between the
electron spin and the nuclear spin of a fermionic isotopes of Yb, given by the
second term in Eq.\ (\ref{Vcpl_R}). We can separate the corresponding operator
$\hat{\Omega}_\textrm{II}=\hat{i}_\textrm{Yb}\cdot \hat{s}$ into three
components,
\begin{align}
\hat{\Omega}_\textrm{II} &= \hat{\Omega}_\textrm{II}^0 +
\hat{\Omega}_\textrm{II}^{+1} + \hat{\Omega}_\textrm{II}^{-1} \nonumber \\
&= \hat{i}_{\textrm{Yb},z}\hat{s}_z+\frac{1}{2}
\hat{i}_{\textrm{Yb}-}\hat{s}_+ + \frac{1}{2} \hat{i}_{\textrm{Yb}+}\hat{s}_-,
\label{eq:OmegaII}
\end{align}
where $\hat{s}_\pm$ and $\hat{i}_{\textrm{Yb}\pm}$ are raising and lowering
operators. The superscripts on the components $\hat{\Omega}_\textrm{II}^x$
correspond to the selection rule $\Delta m_f=0$, $\pm1$, and we will similarly
refer to mechanisms \textrm{II}$^0$, \textrm{II}$^{+1}$ and \textrm{II}$^{-1}$.
These calculations use $L_{\rm max}=2$ in order to take account of inelastic
decay as discussed in Sec.\ \ref{sec:decay}.

\begin{figure}[tbp]
\includegraphics[width=\columnwidth]{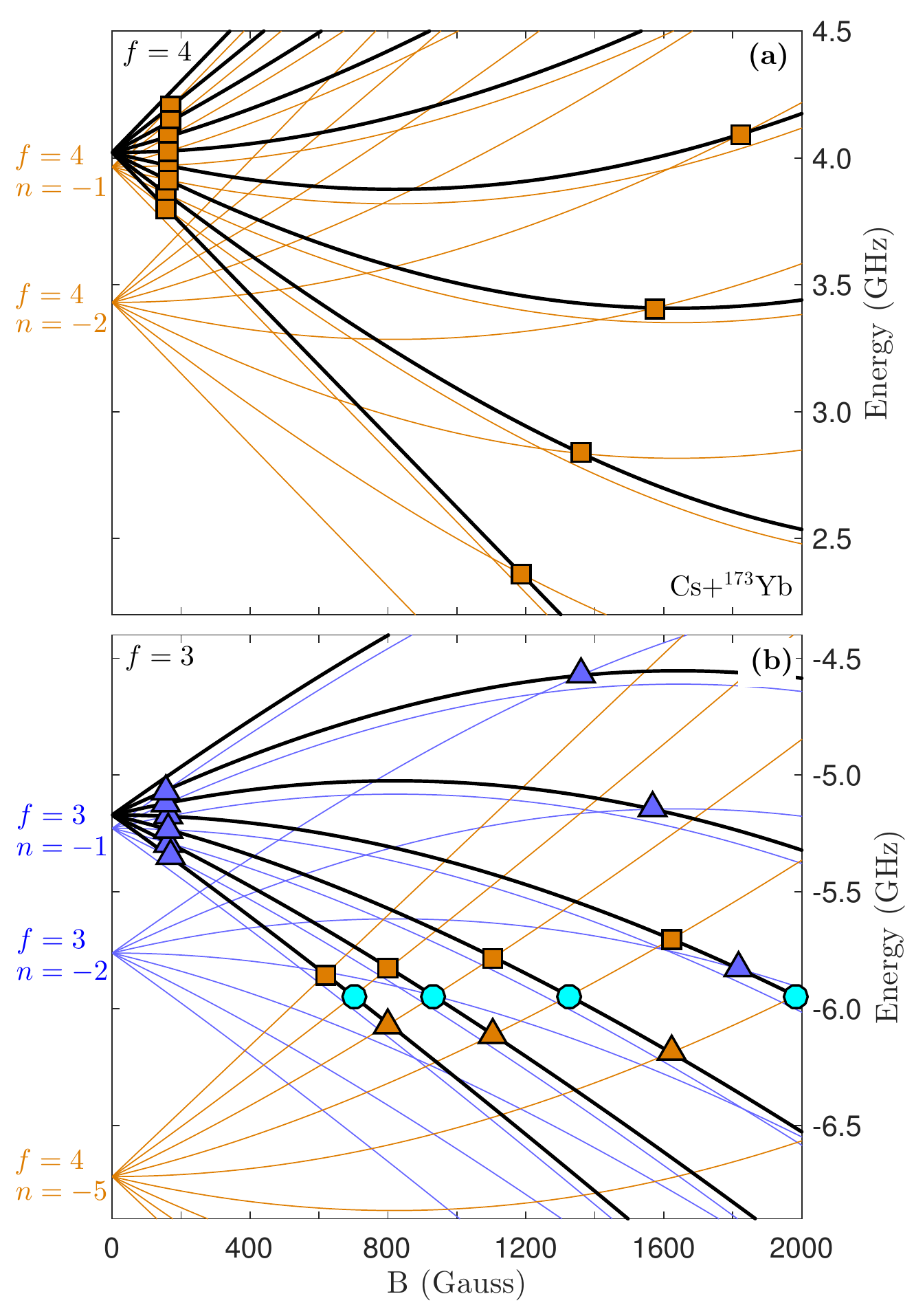}
\caption{
(Color online) Level-crossing diagram with Feshbach resonance positions from
mechanism \textrm{II} for Cs +$^{173}$Yb. The atomic thresholds (thick
black lines) are from the upper ($f=4$) and lower ($f=3$) hyperfine manifolds
in (a) and (b), respectively. The quantum numbers ($f$, $n$) are given on the
left-hand side for each manifold of molecular levels (thin colored lines). The
solid squares, circles and triangles show the positions of Feshbach resonances
caused by mechanism \textrm{II}, with $\Delta m_f=+1$, $0$
and $-1$, respectively.} \label{diagram_II}
\end{figure}

The selection rule on $\Delta m_f$ is less restrictive for mechanism II than
for mechanism I, and allows Feshbach resonances with $\Delta f=0$ as well as
$\Delta f=1$. Figure \ref{diagram_II} shows how the bound states cross the
$f=3$ and $f=4$ scattering thresholds for Cs+$^{173}$Yb. We consider resonances
that arise at crossings where there are direct couplings due to mechanism II,
which are shown as circles, squares and triangles for mechanisms
\textrm{II}$^0$, \textrm{II}$^{+1}$ and \textrm{II}$^{-1}$, respectively.

Many more resonances arise than for mechanism I. In particular, there is a set
of resonances at low field, where the thresholds are crossed by the least-bound
state ($n=-1$) with the same $f$ but $\Delta m_f=-1$ for $f=3$ or $\Delta m_f=+1$
for $f=4$. The corresponding resonance positions are approximately
\begin{equation}
\label{Bres_n1}
B_\textrm{res}(n=-1)=\frac{(2i_\textrm{Cs}+1)}{|\Delta m_f| g_\textit{s}\mu_{\rm B}}E_\textrm{b}(n=-1)
\end{equation}
where $E_\textrm{b}(n=-1)$ is the binding energy of the least-bound state at
$B=0$. For Cs+$^{173}$Yb, Eq.\ (\ref{Bres_n1}) gives
$B_\textrm{res}(n=-1)=163$~G, consistent with the crossings shown in Fig.\
\ref{diagram_II}. The deviations from Eq.\ (\ref{Bres_n1}) are at most a few G
and arise principally from the non-linearity of the atomic Zeeman effect. The
resonance position from the least-bound state with $\Delta f=0$ is
approximately the same in the $f=3$ and $f=4$ manifolds. Even for a system
where the binding energy is unknown, the least-bound state is always within
$36\hbar^2/(2\mu \bar{a}^2)$ of threshold \cite{Gao:2000}, and resonances of
this type exist provided $f$ remains a nearly good quantum number at fields up
to $B_\textrm{res}(n=-1)$; this is the case for Cs or Rb interacting with
either Yb or Sr.

There are also resonances where bound states with $n=-2$ cross thresholds with
the same $f$. These start around $B=1200$~G, but are much more spread out in
field than those for $n=-1$ because the atomic Zeeman effect is nonlinear at
higher fields.

\begin{figure}[tbp]
\includegraphics[width=\columnwidth]{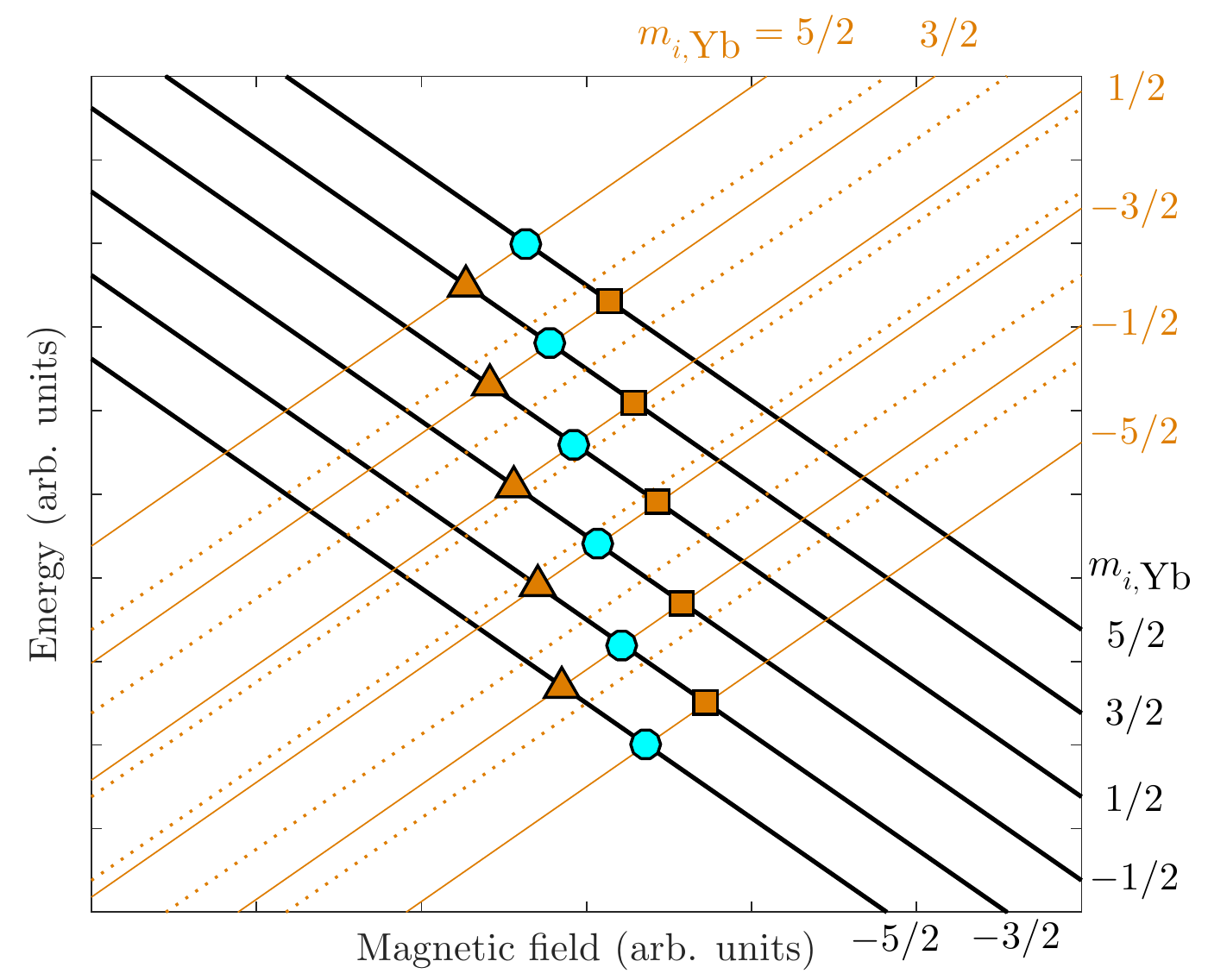}
\caption{
(Color online) Schematic diagram demonstrating the splitting pattern for a set
of resonances arising from a single crossing in Fig.\ \ref{diagram_II}, caused
by mechanism \textrm{II} for Cs+$^{173}$Yb. The atomic thresholds (thick black
lines) and molecular bound states (thin brown lines) are labeled by
$m_{i,\textrm{Yb}}$. The solid squares, circles and triangles indicate the
resonance positions for sets with $\Delta m_f=+1$, $0$ and $-1$, respectively;
only one set appears at each crossing in Fig.\ \ref{diagram_II}. The dotted
lines show the bound states without shifts due to mechanism II.
The energy spacings are typically less that 1~MHz and the set of resonances
typically spans less than 1~G.}
\label{LC_splitting_II}
\end{figure}

Each crossing point in Fig.\ \ref{diagram_II} gives rise to a set of closely
spaced resonances due to states with different $m_{i,\textrm{Yb}}$, as shown
schematically in Fig.\ \ref{LC_splitting_II}. Resonances for different
$m_{i,\textrm{Yb}}$ have different widths, as discussed below. The selection
rule on the nuclear spin projection is $\Delta m_{i,\textrm{Yb}}=-\Delta m_f$;
thus Feshbach resonances occur at different crossing points in the pattern for
mechanisms \textrm{II}$^0$, \textrm{II}$^{+1}$ and \textrm{II}$^{-1}$,
indicated by circles, squares and triangles respectively. The splitting of the
threshold levels is determined solely by the Yb nuclear Zeeman term in Eq.\
\eqref{Yb_Zeeman}, while the splitting of the molecular levels has an
additional contribution from the diagonal matrix elements associated with
mechanism II. Without this additional contribution, all the resonances for the
same value of $\Delta m_f$ would occur at the same field, but its presence
separates the resonances for different $m_{i,\textrm{Yb}}$.

\begin{figure}[tbp]
\includegraphics[width=\columnwidth]{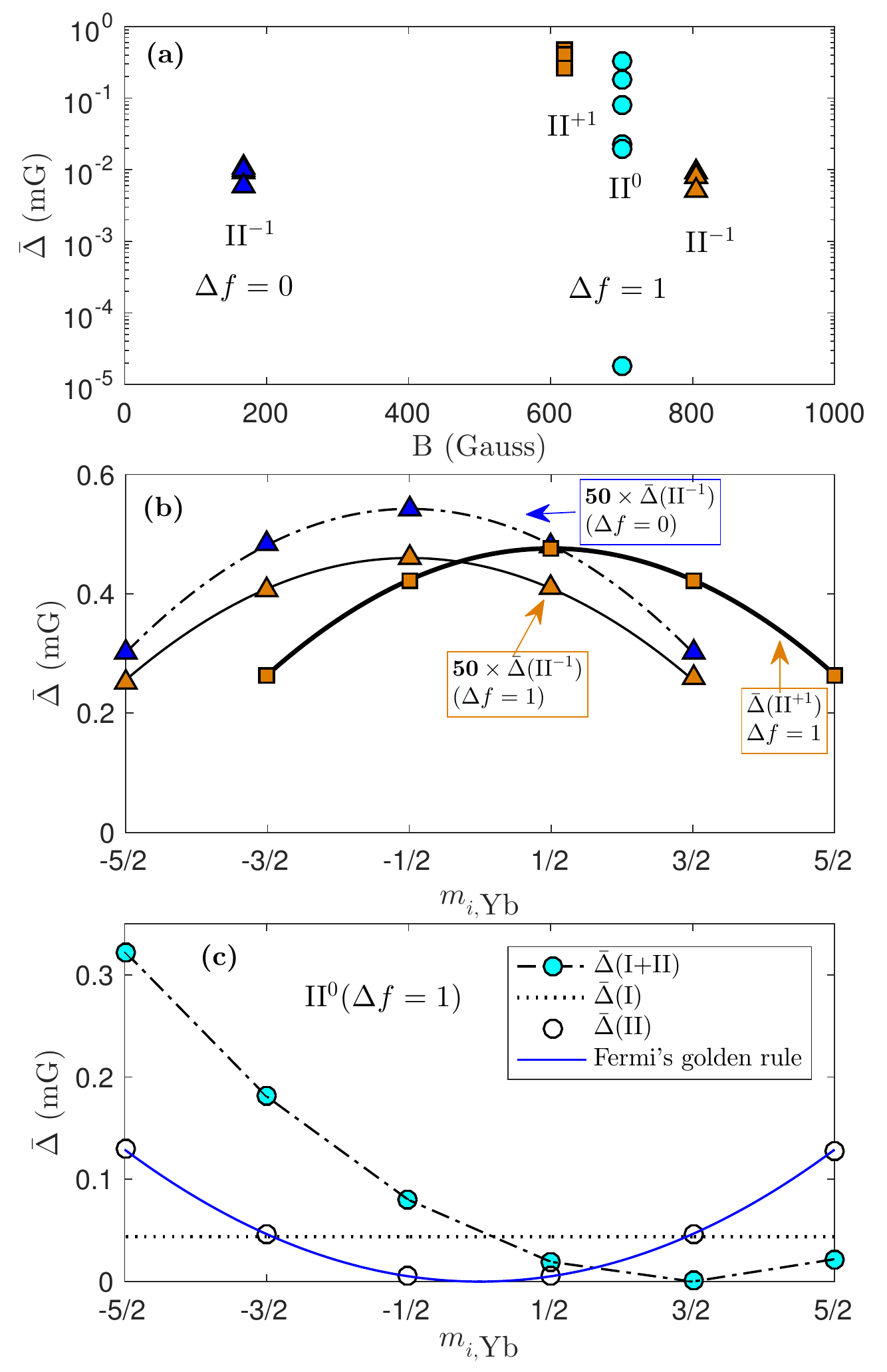}
\caption{(Color online) Widths of the resonances caused by mechanism II for
Cs+$^{173}$Yb. (a) Overview of the sets of resonances arising from the four
crossings at the $f=3$, $m_f=3$ threshold in Fig.\ \ref{diagram_II}. (b) Widths
within each set with $\Delta m_f=\pm1$ as a function of $m_{i,\textrm{Yb}}$.
The lines connecting the points are parabolas as expected from Fermi's golden
rule. (c) Widths within the set with $\Delta m_f=0$, showing separate
contributions from mechanisms I and II and their combination. Symbols
correspond to those in  Fig.\ \ref{diagram_II}.} \label{Mech_II_width}
\end{figure}

General properties of the widths of the resonances can be inferred from Fermi's
golden rule. By contrast with mechanism I, the spin factor in the resonance
widths, $\langle\alpha'|\hat{\Omega}_\textrm{II}|\alpha\rangle^2$, does not
fall to zero as $B\to 0$. This might seem to suggest usefully large widths for
the $\Delta f=0$ resonances that are guaranteed to exist at low field. However,
the radial contribution to the resonance widths, $\langle
n|\omega_\textrm{II}(R)|k\rangle^2$, is proportional to $E_\textrm{b}^{2/3}$
\cite{Alk1S_2013PRA_RbYb_CsYb} where $E_\textrm{b}$ is the binding energy of
the resonant state below the threshold that supports it; through Eq.\
\eqref{Bres_n1}, the width is thus proportional to $B_\textrm{res}^{2/3}$.
Thus, although low-field $\Delta f=0$ resonances arising from mechanism II are
guaranteed to exist, their widths are also somewhat suppressed, albeit more
weakly and for different reasons than for those arising from mechanism I.

There are also $\Delta f=1$ resonances from mechanism II at the $f=3$
thresholds. At each threshold, these occur in three sets, corresponding to the
three allowed values of $\Delta m_f$.
As for the resonances arising from mechanism I, which also have $\Delta f=1$,
these resonances exist at low fields only if the binding energies are
favorable. As shown in Fig.\ \ref{diagram_II}, they exist for Cs$^{173}$Yb at
the lowest ($m_f=3$) threshold at fields from about 600 G upwards, and from
progressively higher fields at excited thresholds.

\begin{figure}[tbp]
\includegraphics[width=\columnwidth]{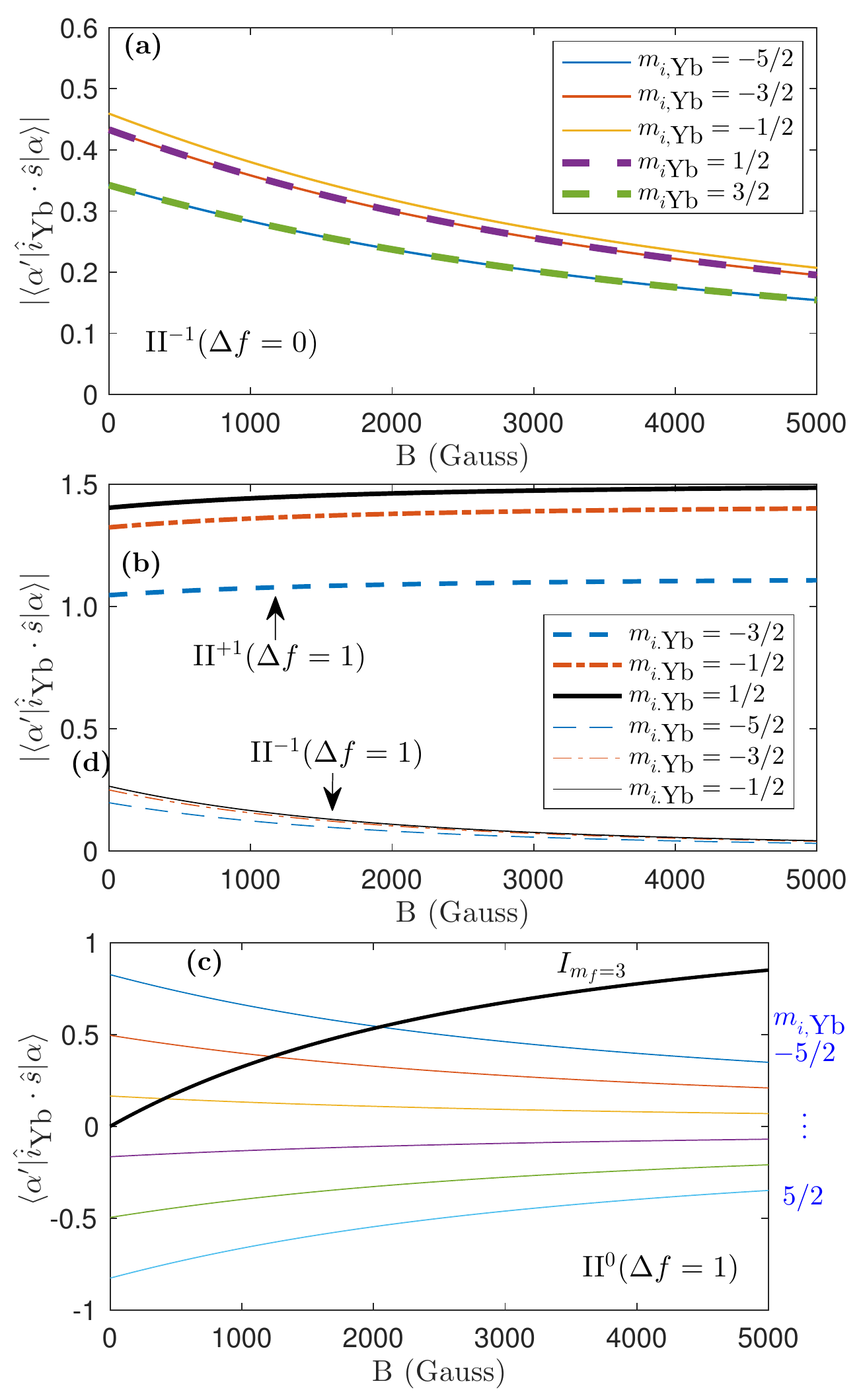}
\caption{(Color online) Spin components $\langle\alpha'|\hat{i}_\textrm{Yb}
\cdot \hat{s}|\alpha\rangle$ of coupling matrix elements for mechanism II. (a)
Absolute values of matrix elements for resonances due to bound states with
$f=3$, $m_f=2$ at the $f=3$, $m_f=3$ threshold, as a function of field, for
different $m_{i,\textrm{Yb}}$. (b) Absolute values of matrix elements for
resonances due to bound states with $f=4$, $m_f=2$ or 4 at the $f=3$, $m_f=3$
threshold. (c) Matrix elements for resonances due to bound states with $f=4$,
$m_f=3$ at the $f=3$, $m_f=3$ threshold. The matrix element for mechanism I is
shown as a thick black line; it adds constructively for $m_{i,\textrm{Yb}}<0$
and destructively for $m_{i,\textrm{Yb}}>0$.} \label{Mech_II_couple}
\end{figure}

The four sets of resonances at the $m_f=3$ thresholds are examined in Fig.\
\ref{Mech_II_width}; three sets have $\Delta f=1$ and one has $\Delta f=0$. The
normalized resonance widths are shown as a function of their resonance
positions in Fig.\ \ref{Mech_II_width}(a) and as a function of
$m_{i,\textrm{Yb}}$ in Fig.\ \ref{Mech_II_width}(b) and (c). For the
II$^{\pm1}$ resonances shown in Fig.\ \ref{Mech_II_width}(b), the resonance
widths are proportional to
$[i_\textrm{Yb}(i_\textrm{Yb}+1)-m_{i,\textrm{Yb}}(m_{i,\textrm{Yb}}\mp1)]$;
this arises simply from the factors due to the lowering/raising operators
$\hat{i}_{\textrm{Yb}\mp}$ in $\hat{\Omega}^{\pm1}$ \cite{Alk1S_2012PRL_LiYb}.
For the \textrm{II}$^{0}$ resonances, the pattern of widths is more complicated
because the atomic scattering and molecular bound states are coupled by both
mechanism \textrm{I} and \textrm{II}. The resonance widths as a function of
$m_{i,\textrm{Yb}}$ are shown in Fig.\ \ref{Mech_II_width}(c) for each
mechanism separately and for the combination. For mechanism I alone, the Yb
nuclear spin is not involved, so the width is constant at
$\bar{\Delta}=$0.04~mG. For mechanism II alone, the width is proportional to
$m_{i,\textrm{Yb}}^2$ due to the operator $\hat{i}_\textrm{Yb,z}$ in
$\hat{\Omega}_\textrm{II}^0$. However, the actual resonance width
$\bar{\Delta}(\textrm{I}+\textrm{II})$ is proportional to the square of the sum
of the coupling matrix elements. This increases the widths for negative
$m_{i,\textrm{Yb}}$ and reduces those for positive $m_{i,\textrm{Yb}}$.

The relative strengths of different sets depend strongly on the electron-spin
components of the states that are coupled. For example, for the II$^{-1}$ set
near 200~G, the spin-dependent matrix element
$\langle\alpha'|\hat{i}_\textrm{Yb} \cdot \hat{s}|\alpha\rangle$ is shown as a
function of magnetic field in Fig.\ \ref{Mech_II_couple}(a). For this
resonance, the dominant electron-spin component is $|m_s=-1/2\rangle$ in both
the scattering and bound states, but the resonance coupling is actually between
$|m_s=1/2\rangle_\textrm{scat}$ and $|m_s=-1/2\rangle_\textrm{bound}$. The
small proportion of $m_s=1/2$ in the scattering state limits the coupling and
so the final resonance width. This component vanishes at high field, so the
matrix elements $\langle\alpha'|\hat{i}_\textrm{Yb} \cdot
\hat{s}|\alpha\rangle$ in Fig.\ \ref{Mech_II_couple}(b) approach zero.

A similar argument applies to the sets of resonances with $\Delta f=1$ and
$\Delta m_f=\pm1$. The corresponding spin-dependent matrix elements are shown
in Fig.\ \ref{Mech_II_couple}(b). For both these sets, the dominant
electron-spin component is $|m_s=1/2\rangle$ in the scattering state and
$|m_s=-1/2\rangle$ in the bound state. Mechanism \textrm{II}$^{+1}$ couples
these two dominant spin components, but \textrm{II}$^{-1}$ couples the smaller
components that vanish at high field. Consequently, the \textrm{II}$^{+1}$
resonances have much larger widths than the \textrm{II}$^{-1}$ resonances, as
shown in Fig.\ \ref{Mech_II_width}(b).

\subsection{Mechanism \textrm{III}}
\label{sec:Mech_III}

Mechanism III is due to the tensor, or anisotropic, hyperfine coupling on the
Yb nucleus, described by the third term in Eq.\ (\ref{Vcpl_R}). Like mechanism
II, it exists only for the fermionic isotopes of Yb. Unlike mechanisms I and
II, this anisotropic coupling can change the rotation of the molecule, with
selection rule $\Delta L=2$; there are also $\Delta L=0$ terms due to this
term, but not for $L=0$. Resonances in s-wave scattering arising from direct
coupling due to mechanism III must therefore come from states with $L=2$. The
other selection rules are $\Delta m_f = 0, \pm 1$ and $\Delta m_{i,\textrm{Yb}}
= 0, \pm 1$. By contrast with mechanism II, a change in $\Delta m_f + \Delta
m_{i,\textrm{Yb}}$ may be compensated by $\Delta M_L\ne0$ to conserve
$M_\textrm{tot}$. The explicit form of the spin-coupling operator
$\hat{\Omega}_\textrm{III}$ is more complicated than for mechanism II, but it
can still be separated by analogy with Eq.\ \eqref{eq:OmegaII} into terms
proportional to $\hat{s}_z$, $\hat{s}_+$ and $\hat{s}_-$. We thus subdivide
mechanism III into mechanisms III$^0$, III$^{+1}$ and III$^{-1}$, respectively.
These calculations use $L_{\rm max}=4$ in order to take account of inelastic
decay as discussed in Sec.\ \ref{sec:decay}.

\begin{figure}[tbp]
\includegraphics[width=\columnwidth]{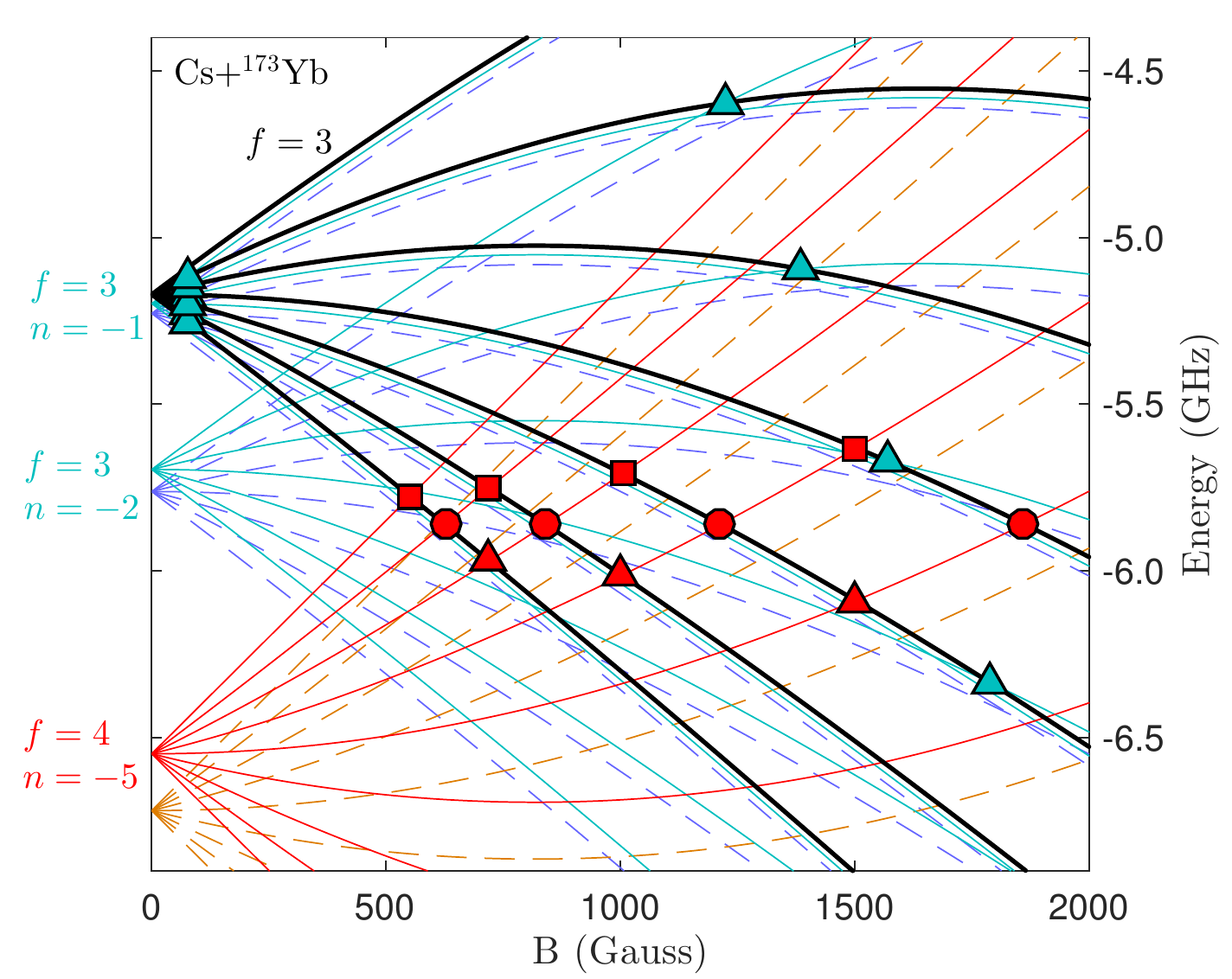}
\caption{
(Color online) Level-crossing diagram with Feshbach resonances from mechanism
\textrm{III} for Cs +$^{173}$Yb. The threshold levels shown (heavy black lines)
are from the lower hyperfine manifolds with $f=3$. The quantum numbers ($f$,
$n$) are labeled on the left-hand side for each manifold of molecular levels
(thin colored lines with solid lines for $L=2$ and dashed lines for $L=0$). The
solid squares, circles and triangles indicate the positions of Feshbach
resonance caused by mechanism \textrm{III}, with $\Delta m_f=+1$, $0$ and $-1$,
respectively.} \label{diagram_III}
\end{figure}

Figure \ref{diagram_III} shows the $L=2$ bound states and the resulting
Feshbach resonances arising from direct coupling due to mechanism III at the
$f=3$ thresholds for Cs+$^{173}$Yb. The $L=0$ bound states are shown as dashed
lines for comparison and are identical to those in Fig.\ \ref{diagram_II}(b).
Each $L=2$ state is immediately above the associated $L=0$ state, with a
spacing proportional to an effective rotational constant, which varies strongly
with the binding energy of the state \footnote{If the least-bound $L=0$ state
is sufficiently close to threshold, the associated $L=2$ state may be above
threshold and quasi-bound or unbound. According to AQDT \cite{Gao:2000}, this
will happen when the background s-wave scattering length is greater than
$\bar{a}$, which is expected in about 50\% of cases.}. This produces a pattern
of resonances very similar to that for mechanism II, but shifted to somewhat
lower field and with additional splittings. Because of the similar separation
of the operator into terms proportional to $\hat{s}_z$, $\hat{s}_+$ and
$\hat{s}_-$, the general conclusions about resonance widths for mechanism II
hold for mechanism III as well.

The most significant difference between mechanisms II and III is in the
internal structure of the sets of resonances. Since the bound states for
mechanism III have $L=2$, there are 5 times as many states, corresponding to
different values of $M_L$. Because of the larger number of states, more
individual crossings within a set can cause Feshbach resonances, and there can
be multiple resonances at each threshold. Within the set of bound states for
each $M_\textrm{tot}$, the states with different $M_L$ and $m_{i,\textrm{Yb}}$
are mixed by coupling due to mechanism III and the Yb quadrupole term. There
are additional small effects due to spin-rotation and Cs quadrupole coupling.
However, the nuclear Zeeman effect and the diagonal matrix elements due to
mechanism II separate the states according to $m_{i,\textrm{Yb}}$, and these
splittings are generally larger than the couplings between them; accordingly,
$m_{i,\textrm{Yb}}$ and $M_L$ remain useful labels, even though they are not
fully conserved.

\begin{figure*}[tbp]
\includegraphics[width=\textwidth]{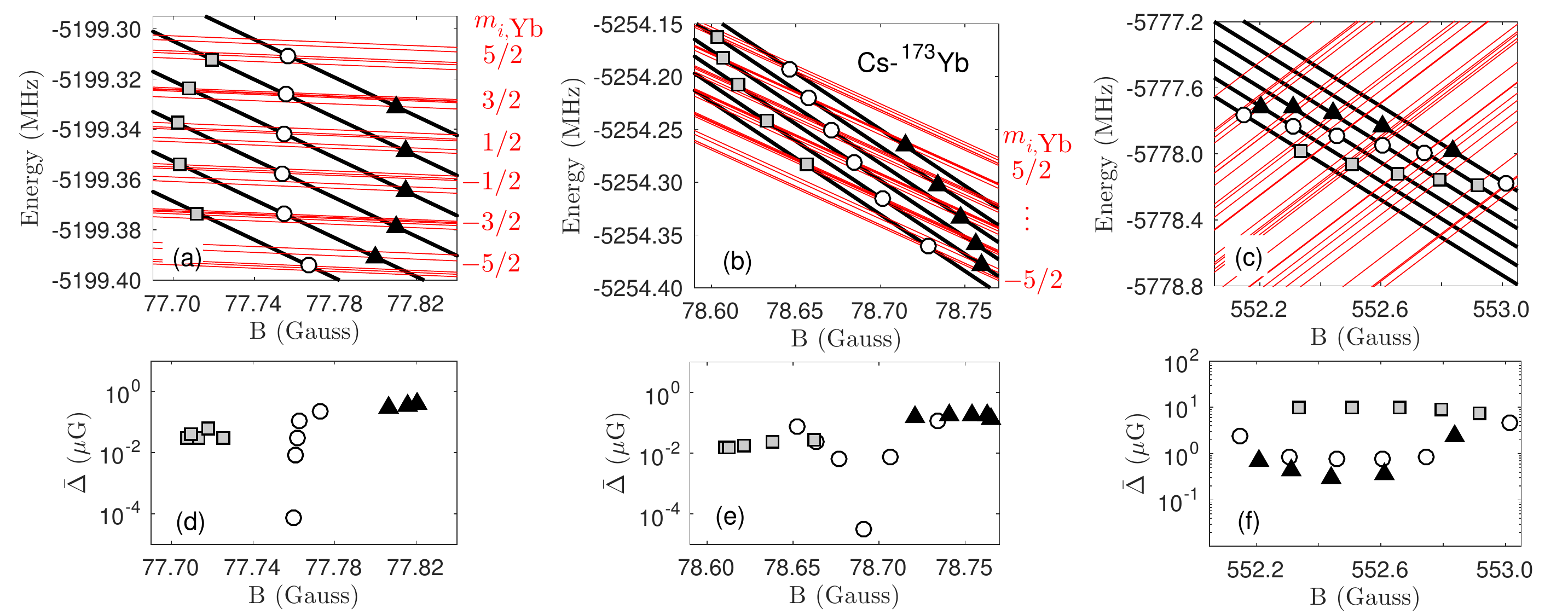}
\caption{(Color online) Structure of sets of resonances caused by mechanism III
for Cs+$^{173}$Yb. Panels (a), (b) and (c) show bound-state levels (thin red
lines) crossing thresholds (thick black lines) for three representative
resonances, with crossings that cause Feshbach resonances due to direct
couplings marked by symbols. Grey squares, white circles, and black triangles
show resonances with $\Delta m_{i,\textrm{Yb}}=+1$, 0, and $-1$, respectively.
(d), (e) and (f) show the corresponding normalized resonance widths
$\bar{\Delta}$ as a function of magnetic field. (a) and (d): set of resonances
near 78~G at the ($f=3$, $m_f=1$) threshold; (b) and (e): set of resonances
near 79~G at the ($f=3$, $m_f=3$) threshold; (c) and (f) set of resonances near
552~G at the ($f=3$, $m_f=3$) threshold.} \label{Mech_III_split}
\end{figure*}

Figure \ref{Mech_III_split} shows the crossing diagrams and the widths of the
resonances in each set as a function of position for three typical examples.
The first example is the set of resonances due to bound states with $f=3$,
$m_f=0$ and $n=-1$ crossing the $f=3$, $m_f=1$ threshold. In this case, the
splitting between the bound states is similar to that between the thresholds.
This is because the diagonal matrix elements of mechanism II and III are both
proportional to the expectation value $\langle m_s \rangle$ of the electron
spin projection, and the bound states have $m_f=0$, for which $\langle m_s
\rangle =0$ at low field. However, states with the same $m_{i,\textrm{Yb}}$ but
different $M_L$ are separated by the Yb quadrupole term. The resulting
resonances are separated into three subsets corresponding to $\Delta m_{i
\textrm{Yb}}=+1,0,-1$, shown in Fig.\ \ref{Mech_III_split} by grey squares,
open circles, and black triangles, respectively. The splitting between the
subsets is governed mostly by the nuclear Zeeman effect and is approximately
$g_\textrm{Yb} \mu_\textrm{B} B / \Delta\mu$. The patterns of widths for
different $m_{i,\textrm{Yb}}$ within these subsets resemble those seen in Fig.\
\ref{Mech_II_width}(b) and (c) for mechanism II, but are distorted by the
mixing of the states.

The second example is a similar set of resonances with $\Delta f=0$, but due to
$f=3$, $m_f=2$ bound states crossing the $f=3$, $m_f=3$ threshold. In this
case, the expectation value $\langle m_s \rangle$ for the bound states is not
near zero, so the molecular states have substantially different splittings to
the thresholds. This separates each of the three subsets (corresponding to
$\Delta m_{i,\textrm{Yb}}=+1,0,-1$) according to $m_{i,\textrm{Yb}}$, such that
the subsets just overlap in field. The widths are shown in Fig.\
\ref{Mech_III_split} and again show the expected patterns with respect to $m_{i
\textrm{Yb}}$. However, the resonances near the middle of this pattern are the
narrowest, so that in loss spectroscopy the resonances would effectively form
\emph{two} groups in field.

The third example is also at the $m_f=3$ threshold, but from a set of
resonances with $\Delta f=1$. The $f=4$ bound states that cause these
resonances are more deeply bound, so the diagonal matrix elements of the
spin-dependent terms are much larger. The bound states can still be labelled by
$m_{i,\textrm{Yb}}$, but the effect of mechanism II is so large that the
ordering of the bound states is reversed from that of the thresholds. The
splittings between states with the same $m_{i,\textrm{Yb}}$ but different
$M_L$, due to mechanism III and the Yb quadrupole term, are also much larger,
such that the multiplets overlap in some cases. The three subsets with
different values of $\Delta m_{i,\textrm{Yb}}$ now completely overlap. As a
result, there is no obvious structure in the pattern of widths as a function of
field.

These three examples qualitatively explain the patterns observed in loss
spectroscopy of similar resonances in $^{87}$Rb+$^{87}$Sr
\cite{Alk1S_2018NP_RbSr}. Figure 1 of Ref.\ \cite{Alk1S_2018NP_RbSr} showed
loss patterns for three different sets of resonances due to mechanism III. One
of these was for a set of resonances due to bound states with $f=1$, $m_f=0$
crossing the $f=1$, $m_f=1$ threshold; these bound states have $\langle m_s
\rangle \sim 0$, so produce a triple peak as in example 1 above. Another was
for a set of resonances due to bound states with $f=1$, $m_f=-1$ crossing the
$f=1$, $m_f=0$ threshold; these bound states have $\langle m_s \rangle \ne 0$,
so produce a double peak as in example 2 above. The third was for a set of
resonances due to bound states with $f=2$, $m_f=-2$ crossing the $f=1$,
$m_f=-1$ threshold; these deeper bound states are more strongly split and mixed
with one another, so produce an unresolved peak as in example 3 above.

\subsection{Inelastic Decay}\label{sec:decay}

Feshbach resonances show signatures of decay when the bound state couples to
inelastic (open) channels below the incoming channel. The primary quantity used
to characterize decay in our calculations is the inelastic decay width
$\Gamma_B^{\textrm{inel}}$. However, the effect of decay on experiments is
better quantified by the lifetime $\tau$ and the resonant scattering length
$a_{\textrm{res}}$. If the magnitude of $a_{\textrm{res}}$ is too small, the
oscillation in the scattering length may not be sufficient to produce
measurable loss in 2-body or 3-body loss spectroscopy; at least
$a_{\textrm{res}} > 100\ a_0$ is probably necessary to produce measurable loss
rates. If the lifetime is too short, molecules formed by magnetoassociation at
the resonance will predissociate before further experimental steps; this may
pose a problem if the lifetimes are milliseconds or less.

For the bosonic isotopes of Yb, mechanism I couples the resonant bound state
only to the incoming channel. For collisions at magnetically excited Cs
thresholds, the Cs quadrupole and tensor hyperfine couplings can in principle
cause decay to inelastic channels with $L=2$, but these terms are small and the
associated decay is very weak. For example, for the resonance near 654~G for Cs
($f=3$, $m_f=-3$) interacting with $^{168}$Yb, we calculate $a_\textrm{res}=3.0
\times 10^{10}\ a_0$, corresponding to $\tau = 2.2 \times 10^5$~s. In the
remainder of this paper, we carry out calculations on bosonic isotopes using
$L_{\rm max}=0$, which suppresses this weak decay.

For the fermionic isotopes of Yb, any of the coupling operators in Eq.\
(\ref{Vcpl_R}) may cause decay, depending on the character of the resonance.
However, there are two situations where there is guaranteed to be very little
decay. The first is for resonances at the lowest Cs hyperfine threshold ($f=3$
and $m_f=3$). These can in principle decay to $L=2$ channels with different
$m_{i,\textrm{Yb}}$, but the associated kinetic energy release is very small
and inelasticity is strongly suppressed by centrifugal barriers in the outgoing
channels.
The second is for II$^{-1}$ resonances at $f=3$ thresholds. For these, the
inelastic channels have $\Delta m_f \ge 1$ relative to the incoming channel,
and thus $\Delta m_f \ge 2$ relative to the bound state. There is no direct
coupling from such bound states to inelastic channels.

For the remaining resonances in fermionic systems, direct decay pathways exist.
Resonances at thresholds with $f=4$ can always decay to $f=3$. This results in
significant decay, with $a_\textrm{res}$ in the range 19 to $125\ a_0$ for the
resonances due to $n=-1$ states, corresponding to $\Gamma_B^\textrm{inel}$ from
$-40$ to $-120\ \mu$G and lifetimes from $3.7$ to $10$~ms. At $f=3$, $m_f<3$
thresholds, resonances due to mechanisms I and II$^{0}$ can decay by mechanisms
II$^{+1}$ and III$^{+1}$, while those due to mechanism II$^{+1}$ can also decay
by mechanisms I, II$^0$ and III$^0$. The resulting decay widths and lifetimes
show considerable variation with $m_{i,\textrm{Yb}}$, but are generally
comparable to those at $f=4$ thresholds; however $a_\textrm{res}$ is
considerably larger because $\bar{\Delta}$ is larger.

For resonances due to mechanism III, the general patterns of decay are similar
to those for resonances due to mechanism II. There are additional decay
pathways to open channels with $L=4$, which sometimes contribute up to 80\% of
the decay widths.

\section{Promising resonances for experimental study} \label{sec:predictions}

In this section we make specific predictions for Feshbach resonances that
appear promising for experimental investigation. We consider resonances in
collisions involving Cs in both its ground state ($f=3$, $m_f=3$) and
magnetically excited states. We highlight the most promising resonances for
each Yb isotope at magnetic fields below 2000~G. Tabulations of resonance
parameters for all resonances at magnetic fields up to $5000$~G are given in
the Supplemental Material.

There are two experimental situations of particular interest. The first is
observation of resonances through their enhancement of collisional processes
such as 3-body recombination, 2-body inelastic loss, or interspecies
thermalization; this is commonly known as Feshbach spectroscopy. The second is
magnetoassociation of pairs of atoms to form weakly bound molecules, which may
be carried out either in an optical trap or in the cells of an optical lattice.

\subsection{Intraspecies Cs collisions}

Any experiment carried out in an optical trap is subject to losses due to
intraspecies as well as interspecies collisions. Even in its ground state
($f=3$, $m_f=3$), ultracold Cs suffers from strong 3-body losses at most
magnetic fields, due to large intraspecies scattering lengths. Similar losses
exist in magnetically and hyperfine excited states, supplemented by 2-body
inelastic losses. The scattering length $a(B)$ was tabulated for the ground
state by Berninger \emph{et al.}\ \cite{Berninger:Cs2:2013}. We have recently
carried out scattering calculations for pairs of excited Cs atoms in the same
state ($f$,$m_f$), using the interaction potentials of Ref.\
\cite{Berninger:Cs2:2013}, for all $f=3$ and $f=4$ states
\cite{Frye:Cs-excited:2019}. We tabulated both the complex scattering length
$a_\textrm{Cs}=\alpha_\textrm{Cs} - i \beta_\textrm{Cs}$ and the rate
coefficient $k_2$ for intraspecies 2-body loss.
We estimate that values of $k_2$ higher than about $10^{-12}$ cm$^3$\,s$^{-1}$
will obscure losses due to interspecies Feshbach resonances.

For experiments in an optical trap, we estimate that intraspecies scattering
lengths larger than about 2000~$a_0$ will produce 3-body losses dominated by
intraspecies collisions. Even for scattering lengths at the upper end of this
range, it will probably be necessary to work with Cs densities below $10^{12}$
cm$^{-3}$ to moderate intraspecies 3-body losses, and with Yb atoms in large
excess so that Cs losses due to resonant interspecies collisions are
competitive.

For each interspecies resonances in the Tables below, we give calculated values
of $\alpha_\textrm{Cs}$, and $k_2$ where it exists, at the resonance position.

\subsection{Experimental considerations}

\subsubsection{Experiments in optical traps}

Optical traps may be used to trap atoms in any internal state and allow
independent control of the applied magnetic field. Although the atomic cloud is
confined to a small volume, there is nevertheless always some variation in the
magnetic field across the sample. This may arise from a magnetic field gradient
used to levitate the atoms, or from other sources such as curvature in the bias
field. There is also inevitably some time-variation of the field, typically on
the order of a few mG. For the narrow resonances predicted in Cs+Yb, it is
likely that only a part of the cloud will be on resonance at any one time. The
resulting loss signal will then be proportional to the range of fields over
which $|a(B)|$ exceeds a critical value $a_{\rm crit}$. For resonances in
elastic scattering, this range is proportional to $a_\textrm{bg}\Delta$. As
described above, for Cs+Yb we have chosen to tabulate the normalized width
$\bar{\Delta}=(a_\textrm{bg}/\bar{a})\Delta$, which retains the dimensions of
field. The narrowest resonance observed in recent experiments on RbSr
\cite{Alk1S_2018NP_RbSr} had a calculated normalized width
$\bar{\Delta}=0.0043$ mG. In this section we tabulate resonances for which
$\bar{\Delta} > 0.04$~mG, except below 200~G, where we tabulate resonances with
$\bar{\Delta} > 0.004$~mG.

Feshbach resonances may also be detected through enhanced interspecies
thermalization \cite{Cho:RbCs:2013}. This is particularly attractive for
Cs+$^{173}$Yb, where the background scattering length is very low and there
will be very little interspecies thermalization away from resonance. The rate
of interspecies thermalization is also expected to be approximately
proportional to $\bar{\Delta}$.

Cs atoms in $f=4$ excited states are predicted to decay quickly by 2-body
inelastic processes \cite{Frye:Cs-excited:2019}. We therefore focus on Cs+Yb
resonances involving Cs atoms in $f=3$ states. For bosonic isotopes of Yb at
any threshold, and for fermionic isotopes at the lowest threshold, the
interspecies resonances are undecayed and the scattering length passes through
a pole at resonance. However, for fermionic isotopes at thresholds with $m_f<3$
the resonances may be decayed, as described in Sec.\ \ref{sec:decay}; the pole
is then replaced by a more complicated lineshape, which for the resonances
considered here is an essentially symmetric oscillation of amplitude $\pm
a_{\rm res}/2$. If $a_{\rm res}$ is less than about 100~$a_0$, an interspecies
resonance may not produce a significant peak in 3-body loss.

Interspecies 2-body loss occurs only with fermionic isotopes of Yb in
combination with excited states of Cs. It is very weak away from resonance, but
shows a narrow peak of height proportional to $a_{\rm res}$ at resonance. There
may be some resonances and conditions under which interspecies 2-body loss is
faster than 3-body loss.

\subsubsection{Experiments in lattices}

Experiments in 3D optical lattices have several advantages. By loading
quantum-degenerate gases into the lattice and exploiting the
superfluid-to-Mott-insulator transition \cite{Greiner:2002}, the number of
atoms loaded onto a lattice site can be controlled and tunneling suppressed.
Under such conditions, intraspecies losses can be completely eliminated.
Experiments may thus be performed with any internal state and at any magnetic
field, without restriction on the intraspecies scattering properties; this is
particularly beneficial when working with atoms such as Cs, where intraspecies
loss may otherwise be a limiting factor. The use of an optical lattice also
removes the need for a field gradient to levitate the cloud against gravity.

Experiments in lattices are still subject to interspecies 2-body loss when it
is present. For fermionic isotopes of Yb, combined with excited states of Cs,
it may be possible to detect resonances by searching for 2-body loss as a
function of magnetic field in a lattice.

\subsubsection{Magnetoassociation}

Magnetoassociation may be carried out either in an optical trap or in a lattice
cell containing one atom of each type. In a confined system, the scattering
continuum above threshold is replaced by a series of quantized translational
levels. A scattering resonance then appears as a series of avoided crossings
between the molecular states and these quantized levels. The strengths (energy
widths) of the avoided crossings are proportional to
$(a_\textrm{bg}\Delta)^{1/2}$ \cite{Mies:Feshbach:2000, Julienne:2004}. In
magnetoassociation, the goal is to sweep the magnetic field across the lowest
of the avoided crossings slowly enough to achieve adiabatic passage. The
maximum sweep speed that achieves this is proportional to the square of the
strength and thus to $a_\textrm{bg}\Delta$ \cite{Mies:Feshbach:2000,
Julienne:2004}. Because of this, $\bar{\Delta}$ is an appropriate measure of
the resonance width for magnetoassociation as well as for loss spectroscopy.

A lattice cell confines a pair of atoms more tightly than an optical trap,
increasing the strength of the avoided crossing available for
magnetoassociation. The strength is proportional to $\omega^{3/4}$
\cite{Mies:Feshbach:2000, Julienne:2004}, where $\omega$ is the harmonic trap
frequency \footnote{Where the two species have different trap frequencies, the
relevant frequency is that of the confinement in the relative motion,
$\omega_\textrm{rel}=\sqrt{\omega_A^2\mu/m_A+\omega_B^2\mu/m_B}$
\cite{Deuretzbacher:2008}.}; the maximum speed of the field sweep is thus
proportional to $\omega^{3/2}$.

For a broad resonance, it is relatively easy to sweep the field slowly enough
to achieve adiabatic passage. However, for narrow resonances such as those
considered here, it is more challenging. Field inhomogeneity results only in
different parts of the sample crossing the resonance at different times. Field
noise, however, may result in repeated crossing and recrossing at speeds that
cause nonadiabatic transitions and loss. Very narrow resonances thus require
very stable fields.

\subsubsection{Molecular lifetimes}

Molecules formed by magnetoassociation at a decayed resonance may themselves
decay (predissociate) spontaneously with lifetime $\tau$, as described in
section \ref{scattering_length}. In practical terms, it is necessary to
stabilize the magnetic field after magnetoassociation sweep before transferring the
molecules to another state. This is likely to be difficult if the molecular
lifetime is less than about 100~$\mu$s.

\subsection{Cs + bosonic Yb}

\begin{table}[tbp]
  \centering
\caption{Experimentally promising resonances in ultracold collisions between Cs
and bosonic isotopes of Yb.}\label{resonances_bosonic}
\begin{ruledtabular}
\begin{tabular}{crrrrr}
\rule{0pt}{2ex} &               & $B_\textrm{res}$       &  $\bar{\Delta} $
                    &  $\alpha_\textrm{Cs}$  &    $k_2$ \\
Cs-Yb          & $m_f$    &      (G)                                &  (mG)
& ($a_0$)   & (cm$^3$ s$^{-1}$) \\ \hline
 133-168	& $-3$	& 654		& $-0.074$	& $1.59\times10^{3}$	& $2.25\times10^{-13}$  \\
       \hline
 133-170 & $-1$	& 366		& $-0.098$	& $3.53\times10^{3}$	& $2.41\times10^{-11}$  \\
		& $-1$	& 1273		& 1.2 	& $1.99\times10^{2}$	& $9.62\times10^{-11}$  \\
  \hline
 133-172	& 2	& 602		& 0.051	& $1.91\times10^{3}$	& $1.75\times10^{-11}$  \\
		& 1	& 914		& 0.19	& $-1.89\times10^{5}$	& $9.70\times10^{-12}$  \\
		& 0	& 1528		& 0.64	& $7.10\times10^{2}$	& $1.76\times10^{-13}$  \\
\hline
133-174	& 3	& 964		& 0.17	& $1.11\times10^{3}$	&   \\
		& 2	& 1252		& 0.60	& $1.21\times10^{3}$	& $2.91\times10^{-14}$  \\
		& 1	& 1699		& 1.6		& $1.26\times10^{3}$	& $2.15\times10^{-13}$  \\
\hline
133-176	& 3	& 1497		& 5.9   	& $2.43\times10^{3}$	&   \\
		& 2	& 1866		& 18		& $2.20\times10^{3}$	& $1.33\times10^{-13}$  \\
		& $-3$	& 1559		& $-35$    	& $1.16\times10^{3}$	& $1.22\times10^{-13}$  \\
		& $-3$	& 3359		& 160	& $2.31\times10^{2}$	& $4.79\times10^{-14}$  \\
\end{tabular}
\end{ruledtabular}
\end{table}

For Cs interacting with bosonic isotopes of Yb, there are only a few resonances
located below $2000$~G. These are all caused by mechanism I. Inelastic decay is
negligible for these resonances, even for excited states of Cs, as discussed in
Section \ref{sec:decay}. The important properties are the resonance position
and width, as well as the properties relevant to background loss of Cs for
experiments in an optical trap. Table \ref{resonances_bosonic} lists all
resonances that meet the width criteria described above, together with some
additional ones that warrant discussion.

The resonances for $^{176}$Yb are the strongest in Table
\ref{resonances_bosonic}, and also have small two-body loss rates for Cs. The
pair of resonances near $1559$~G and $3359$~G are from a double crossing
between the atomic and molecular states with $m_f=-3$. The relatively large
normalized widths $\bar{\Delta}$ for these occur both because the background
scattering length is large (798~$a_0$) and because the difference between the
magnetic moments of the atomic and molecular states is small near such a double
crossing \cite{Alk1S_2013PRA_RbYb_CsYb}. The resonance at $3359$~G is included
in Table \ref{resonances_bosonic}, despite its high field, because it is
unusually wide and is also in a field range where 3-body loss of Cs is expected
to be relatively slow. These resonances are promising for loss spectroscopy.
However, $^{176}$Yb has a small negative intraspecies scattering length
\cite{Kitagawa:2008}, which leads to collapse of its condensates
\cite{Fukuhara:2009}, so that a lattice with a high filling fraction will be
hard to produce.

The Yb isotopes that are most easily cooled to degeneracy, and are thus most
suitable for formation of Mott insulators, are $^{174}$Yb
\cite{Fukuhara:Mott:2009} and $^{170}$Yb \cite{Sugawa:Mott:2011}. The
normalized widths of the resonances for these isotopes are smaller than for
$^{176}$Yb, but magnetoassociation in an optical lattice may still be feasible.
For $^{170}$Yb, two-body loss of Cs atoms may prevent observation of the
resonances by loss spectroscopy. The three resonances for $^{174}$Yb appear
more suitable for loss spectroscopy, though 3-body losses of Cs atoms are
expected to be fairly fast.

$^{172}$Yb has a large negative intraspecies scattering length
\cite{Kitagawa:2008}, and has not been cooled to degeneracy. It nevertheless
has resonances that may be observable by loss spectroscopy. The resonance near
$1528$~G for $^{172}$Yb with Cs ($f=3$, $m_f=0$) appears particularly suitable
for this because of the relatively small background losses expected for Cs
atoms.

$^{168}$Yb has a very low isotopic abundance and the only resonance available
below 2000~G is the one near $654$~G. This resonance might be observable by
loss spectroscopy, but has no obvious advantages over those for more abundant
isotopes.

\subsection{Cs + fermionic Yb}

\begin{table*}[tbp]
\caption{Experimentally promising resonances in ultracold collisions between Cs
and $^{171}$Yb. Parameters are given for the widest resonance in each set, and
the corresponding value of $m_{i,\textrm{Yb}}$ is given.
  }\label{resonances_171}
\begin{ruledtabular}
\begin{tabular}{crccrrrrrr}
\rule{0pt}{2ex}  		&   &	  &	& $B_\textrm{res}$ 	& $\bar{\Delta}$	& $a_{\textrm{res}}$	& $\tau$	& $\alpha_\textrm{Cs}$	& $k_2$			\\
                   $(f, m_f)$  & $m_{i, \textrm{Yb}}$  & $(\Delta f, \Delta m_f)$	& mechanism	& (G)			& (mG)			& ($a_0$)			& (s)		& ($a_0$)				& (cm$^3$ s$^{-1}$)  \\
 \hline
                   $(3, 3)$  & $-1/2$  & $(0, -1)$ 	&	II	&		75     	& 0.0048	& $1.7\times10^{8}$	& $9.7\times10^{4}$ 	& $1.36\times10^{3}$	&          \\
 $(3, 3)$  & $-1/2$  & $(1, 1)$	&	III	&		81		& 0.0063 & $6.0\times10^{5}$ & 37      &  $1.42\times10^{3}$      &             \\
                  $(3, 3)$  & $1/2$  & $(1, 1)$	&	II	&		149   	& 0.33	& $1.1\times10^{12}$	& $1.2\times10^{6}$ 	& $1.82\times10^{3}$	&            \\
                  $(3, 3)$  & $1/2$  & $(1, 0)$	&	I+II	&		171   	& 0.065	& $4.1\times10^{11}$	& $2.8\times10^{6}$ 	& $1.91\times10^{3}$	&           \\
                  $(3, 2)$  & $-1/2$  & $(0, -1)$	&	II	&		74     	& 0.0081	& $1.6\times10^{8}$	& $5.6\times10^{4}$ 	& $2.08\times10^{3}$	& $7.36\times10^{-13}$            \\
 $(3, 2)$  & $-1/2$  & $(1, 1)$   & III & 113 & 0.0065 & 9.6 & $7.7\times10^{-4}$      &     $-3.44\times10^{3}$	  &    $1.53\times10^{-12}$         \\
                  $(3, 2)$  & $1/2$  & $(1, 1)$	&	II	&		203   	& 0.34	& $9.5\times10^{2}$	& $1.4\times10^{-3}$  	& $7.66\times10^{2}$	& $3.35\times10^{-14}$            \\
                  $(3, 2)$  & $1/2$  & $(1, 0)$	&	I+II	&		247   	& 0.18	& $9.9\times10^{2}$	& $3.4\times10^{-3}$  	& $1.05\times10^{3}$	& $6.57\times10^{-14}$            \\
                  $(3, 2)$  & $-1/2$  & $(1, -1)$	&	II	&		315   	& 0.054	& $6.3\times10^{6}$	& 87  	& $1.36\times10^{3}$	& $9.82\times10^{-14}$            \\
                  $(3, 2)$  & $-1/2$  & $(0, -1)$	&	II	&		1517  	& 0.057	& $2.1\times10^{5}$	& 17  	& $1.64\times10^{3}$	& $5.55\times10^{-14}$            \\
                  $(3, 1)$  & $-1/2$  & $(0, -1)$	& 	II	&		74     	& 0.0097	& $1.7\times10^{8}$	& $4.9\times10^{4}$ 	& $3.64\times10^{2}$	& $1.44\times10^{-12}$            \\
 $(3, 1)$  & $-1/2$   & $(1, 1)$  & III & 180 & 0.0074 & 3.3 & $3.6\times10^{-4}$      &      $-2.47\times10^{3}$   &	$8.27\times10^{-12}$           \\
                  $(3, 1)$  & $1/2$  & $(1, 1)$	& 	II	&		315	        & 0.38	& $6.0\times10^{2}$	& $1.2\times10^{-3}$  	& $-2.20\times10^{3}$	& $1.88\times10^{-12}$            \\
                  $(3, 1)$  & $1/2$  & $(1, 0)$	& 	I+II	&		423   	& 0.44	& $6.2\times10^{2}$	& $1.3\times10^{-3}$  	& $5.39\times10^{2}$	& $3.17\times10^{-14}$            \\
                  $(3, 1)$  & $-1/2$  & $(1, -1)$	& 	II	&		613   	& 0.14	& $1.4\times10^{6}$	& 11  	& $1.18\times10^{3}$	& $1.49\times10^{-11}$            \\
                  $(3, 1)$  & $-1/2$  & $(0, -1)$	& 	II	&		1373  	& 0.069	& $2.5\times10^{5}$	& 14  	& $-1.48\times10^{2}$	& $2.24\times10^{-11}$           \\
                  $(3, 0)$  & $-1/2$  & $(0, -1)$        &	II	&		73     	& 0.0097	& $1.8\times10^{8}$	& $5.0\times10^{4}$ 	& $-2.44\times10^{3}$	& $2.14\times10^{-11}$           \\
                  $(3, 0)$  & $1/2$  & $(1, 1)$        &	II	&		613   	& 0.46	& $1.4\times10^{3}$  & $3.3\times10^{-3}$  	& $-2.03\times10^{3}$	& $1.01\times10^{-11}$           \\
                  $(3, 0)$  & $1/2$  & $(1, 0)$        &	I+II	&		934   	& 1.22	& $9.1\times10^{2}$	& $9.2\times10^{-4}$  	& $6.77\times10^{2}$	& $9.23\times10^{-12}$            \\
                  $(3, 0)$  & $-1/2$  & $(0, -1)$        &	II	&		1243 	& 0.067	& $3.0\times10^{5}$	& 14  	& $1.33\times10^{3}$	& $5.67\times10^{-13}$            \\
                  $(3, 0)$  & $-1/2$   & $(1, -1)$       &	II	&		1444  	& 0.14	& $2.3\times10^{5}$	& 1.9 	& $1.49\times10^{2}$	& $1.53\times10^{-11}$           \\
                  $(3, -1)$  & $-1/2$  & $(0, -1)$       &	II	&		73	        & 0.0081	& $1.9\times10^{8}$	& $6.1\times10^{4}$ 	& $-8.35\times10^{3}$	& $5.31\times10^{-11}$            \\
                  $(3, -1)$  & $-1/2$  & $(0, -1)$       &	II	&		1125 	& 0.056	& $3.6\times10^{5}$	& 16  	& $3.15\times10^{3}$	& $2.79\times10^{-11}$            \\
                  $(3, -1)$   & $1/2$  & $(1, 1)$      &	II	&		1444 	& 0.47	& $3.6\times10^{2}$	& $8.7\times10^{-4}$  	& $8.12\times10^{3}$	& $2.43\times10^{-11}$            \\
                  $(3, -2)$  & $-1/2$  & $(0, -1)$       &	II	&		73     	& 0.0049	& $1.9\times10^{8}$	& $1.0\times10^{5}$ 	& $7.43\times10^{3}$	& $3.90\times10^{-11}$            \\
$(4, -4)$   & $1/2$  & $(0, 1)$	 &	II	&		72     	        & 0.0066	& 97  	    & $3.9\times10^{-2}$	& $2.95\times10^{3}$	& $2.49\times10^{-11}$            \\
$(4, -4)$   & $1/2$  & $(0, 1)$	 &	II	&			927   	& 0.042	& $9.7\times10^{2}$	    & $3.7\times10^{-2}$	& $2.94\times10^{3}$	& $2.39\times10^{-11}$            \\
$(4, -3)$   & $1/2$  & $(0, 1)$	 &	II	&			73      	& 0.011	& 91          	& $2.1\times10^{-2}$  	& $1.10\times10^{3}$	& $1.92\times10^{-10}$           \\
$(4, -3)$   & $1/2$  & $(0, 1)$	 &	II	&			1021  	& 0.075	& $2.4\times10^{2}$  	& $6.3\times10^{-3}$  	& $8.00\times10^{2}$	& $1.02\times10^{-10}$           \\
$(4, -2)$   & $1/2$  & $(0, 1)$	 &	II	&			73     	& 0.015	& 78	                 & $1.4\times10^{-2}$  	& $7.73\times10^{2}$	& $1.68\times10^{-10}$            \\
$(4, -2)$   & $1/2$  & $(0, 1)$	 &	II	&			1126 	& 0.098	& $1.1\times10^{2}$         & $2.7\times10^{-3}$  	& $6.16\times10^{2}$	& $7.59\times10^{-11}$           \\
$(4, -1)$   & $1/2$  & $(0, 1)$	 &	II	&			74     	& 0.016	& 64  	& $1.1\times10^{-2}$  	& $6.68\times10^{2}$	& $1.54\times10^{-10}$            \\
$(4, -1)$   & $1/2$  & $(0, 1)$	 &	II	&			1243	         & 0.11	& 58  	& $1.6\times10^{-3}$   	& $5.95\times10^{2}$	& $7.29\times10^{-11}$            \\
$(4, 0)$   & $1/2$   & $(0, 1)$ 	 &	II	&			74     	& 0.016	& 50  	& $8.4\times10^{-3}$  	& $6.43\times10^{2}$	& $1.50\times10^{-10}$            \\
$(4, 0)$   & $1/2$  & $(0, 1)$  	 &	II	&			1373 	& 0.11	& 34  	& $1.1\times10^{-3}$  	& $6.46\times10^{2}$	& $8.18\times10^{-11}$           \\
$(4, 1)$   & $1/2$   & $(0, 1)$ 	 &	II	&			74     	& 0.015	& 37  	& $6.9\times10^{-3}$ 	& $6.77\times10^{2}$	& $1.55\times10^{-10}$            \\
$(4, 1)$   & $1/2$   & $(0, 1)$ 	 &	II	&			1517 	& 0.098	& 20  	& $9.2\times10^{-4}$  	& $7.67\times10^{2}$	& $9.95\times10^{-11}$            \\
$(4, 2)$   & $1/2$   & $(0, 1)$ 	 &	II	&			75     	& 0.011	& 24  	& $5.8\times10^{-3}$  	& $7.94\times10^{2}$	& $1.69\times10^{-10}$            \\
$(4, 2)$   & $1/2$   & $(0, 1)$ 	 &	II	&			1673 	& 0.075	& 11  	& $8.0\times10^{-4}$  	& $1.00\times10^{3}$	& $1.23\times10^{-10}$            \\
$(4, 3)$   & $1/2$   & $(0, 1)$ 	 &	II	&			75     	& 0.0066	& 12  	& $5.0\times10^{-3}$  	& $1.13\times10^{3}$	& $1.92\times10^{-10}$            \\
$(4, 3)$   & $1/2$  & $(0, 1)$   	 &	II	&			1840 	& 0.042	& 4.8   	& $7.3\times10^{-4}$  	& $1.51\times10^{3}$	& $1.39\times10^{-10}$           \\
\end{tabular}
\end{ruledtabular}
\end{table*}

\begin{table*}[tbp]

\caption{Experimentally promising resonances in ultracold collisions between Cs
and $^{173}$Yb. Parameters are given for the widest resonance in each set, and
the corresponding value of $m_{i,\textrm{Yb}}$ is given.
  }\label{resonances_173}
\begin{ruledtabular}
\begin{tabular}{crccrrrrrr}
\rule{0pt}{2ex}   &  &   &  	& $B_\textrm{res}$ 	& $\bar{\Delta}$	& $a_{\textrm{res}}$	& $\tau$	& $\alpha_\textrm{Cs}$	& $k_2$			\\
          	$(f, m_f)$  & $m_{i, \textrm{Yb}}$  & $(\Delta f, \Delta m_f)$	& mechanism	& (G)			& (mG)			& ($a_0$)			& (s)		& ($a_0$)				& (cm$^3$ s$^{-1}$)  \\
 \hline
         	$(3, 3)$  & $-1/2$  & $(0, -1)$	&      II	& 167   	& 0.011	& $2.7\times10^{13}$	& $7.1\times10^{9}$       &  $1.90\times10^{3}$      &             \\
	 $(3, 3)$  & $-3/2$  & $(1, 1)$    & 	III	& 553 & 	0.0053         &	$6.0\times10^{8}$&   $2.2\times10^{4}$     &   $-2.02\times10^{3}$     &             \\
                 $(3, 3)$  & $1/2$  & $(1, 1)$	&      II	& 620   	& 0.48	& $3.7\times10^{7}$	& 29       &  $2.92\times10^{3}$      &             \\
                 $(3, 3)$  & $-5/2$  & $(1, 0)$	&      I+II	& 700   	& 0.32	& $1.9\times10^{13}$	& $2.5\times10^{7}$       &  $4.39\times10^{3}$      &             \\
                 $(3, 2)$  & $-1/2$  & $(0, -1)$	&      II	& 165   	& 0.018	& $1.3\times10^{9}$	& $2.0\times10^{5}$       &   $3.55\times10^{2}$	&      $5.00\times10^{-15}$        \\
                 $(3, 2)$  & $1/2$  & $(1, 1)$	&      II	& 804   	& 0.49	& $7.5\times10^{2}$	& $7.1\times10^{-4}$       &   $2.31\times10^{3}$	&  $5.94\times10^{-13}$      \\
                 $(3, 2)$  & $-5/2$  & $(1, 0)$	&      I+II	& 933   	& 0.86	& $6.0\times10^{5}$	& 36       &     $-6.08\times10^{4}$	&     $3.49\times10^{-12}$       \\
                 $(3, 1)$  & $-1/2$  & $(0, -1)$	&      II	& 163   	& 0.022	& $8.5\times10^{8}$	& $1.1\times10^{5}$       &    $-7.60\times10^{3}$  &	$2.66\times10^{-11}$     \\
                 $(3, 1)$  & $1/2$  & $(1, 1)$	&      II	& 1107 	&  0.50	& $2.3\times10^{2}$	& $2.5\times10^{-4}$       &    $1.41\times10^{3}$	  &    $6.63\times10^{-13}$            \\
                 $(3, 1)$  & $-5/2$  & $(1, 0)$	&      I+II	& 1326	&  1.80	& $4.5\times10^{5}$	& $1.5\times10^{-1}$       &    $-1.65\times10^{3}$	   &   $1.32\times10^{-11}$          \\
                 $(3, 1)$  & $-1/2$  & $(1, -1)$	&      II	& 1622  & 	0.048	& $4.2\times10^{8}$	& $5.5\times10^{3}$       &    $1.13\times10^{3}$	  &    $2.31\times10^{-13}$         \\
                 $(3, 0)$  & $-1/2$  & $(0, -1)$	&      II	& 161   & 	0.022	& $7.8\times10^{8}$	& $9.6\times10^{4}$      &    $-9.88\times10^{3}$	&     $1.08\times10^{-10}$        \\
                 $(3, 0)$  & $1/2$  & $(1, 1)$	&      II	& 1624 & 	0.52	& 88  	& $1.1\times10^{-4}$       &     $1.05\times10^{3}$	  &   $3.10\times10^{-13}$          \\
                 $(3, 0)$  & $-1/2$  & $(0, -1)$	&      II	& 1815 & 	0.13	& $3.5\times10^{7}$	& $1.0\times10^{3}$       &   $2.15\times10^{3}$  &	$2.14\times10^{-12}$          \\
                 $(3, 0)$  & $-5/2$  & $(1, 0)$     &    I+II	& 1983 & 	3.30	& $5.7\times10^{5}$	& $1.1\times10^{-1}$       &  $-2.63\times10^{2}$	  &   $2.22\times10^{-10}$       \\
                 $(3, -1)$  & $-1/2$  & $(0, -1)$    & 	II	& 159   &  	0.018	& $8.5\times10^{8}$	& $1.2\times10^{5}$       &   $6.95\times10^{3}$  &	$5.15\times10^{-11}$      \\
                 $(3, -1)$  & $-1/2$  & $(0, -1)$    & 	II	& 1566 & 	0.10	& $8.5\times10^{8}$	& $2.2\times10^{4}$       &    $7.35\times10^{4}$	&    $4.09\times10^{-11}$         \\
                 $(3, -2)$  & $-1/2$  & $(0, -1)$    & 	II	& 157   & 	0.011	& $1.1\times10^{9}$	& $2.6\times10^{5}$       &    $4.65\times10^{3}$  &	$1.43\times10^{-11}$          \\
                 $(3, -2)$  & $-1/2$  & $(0, -1)$    & 	II	& 1357 & 	0.058	& $2.1\times10^{9}$	& $7.0\times10^{4}$      &     $1.33\times10^{3}$	&   $2.46\times10^{-13}$          \\
                 $(3, -3)$  & $-1/2$  & $(1, -1)$    & 	II	& 1744 & -0.30	         & $1.0\times10^{10}$	& $1.8\times10^{4}$        &   $9.43\times10^{2}$   &	$1.34\times10^{-13}$      \\
$(4, -4)$  & $-1/2$  & $(0, 1)$    &	II	&	157   	& 0.015	& 37  	& $6.3\times10^{-3}$      &    $2.95\times10^{3}$  &	$2.50\times10^{-11}$          \\
$(4, -4)$  & $-1/2$  & $(0, 1)$    &	II	&	1188 	& 0.069	& 36  	& $7.2\times10^{-4}$     &     $2.96\times10^{3}$  &	$2.27\times10^{-11}$            \\
$(4, -3)$  & $-1/2$  & $(0, 1)$    &	II	&	158   	& 0.026	& 76  	& $7.6\times10^{-3}$      &      $1.01\times10^{3}$	  &    $1.52\times10^{-10}$          \\
$(4, -3)$  & $-1/2$  & $(0, 1)$   &	II	&	1362 	& 0.13	& 35  	& $5.2\times10^{-4}$      &    $7.37\times10^{2}$ &	$9.39\times10^{-11}$         \\
$(4, -2)$  & $-1/2$  & $(0, 1)$    &	II	&	160   	& 0.033	& $1.1\times10^{2}$  	& $8.5\times10^{-3}$      &     $7.05\times10^{2}$	&     $1.24\times10^{-10}$           \\
$(4, -2)$  & $-1/2$  & $(0, 1)$    &	II	&	1574 	& 0.17	& 27  	& $4.3\times10^{-4}$      &    $6.06\times10^{2}$   &	$7.41\times10^{-11}$          \\
$(4, -1)$  & $-1/2$  & $(0, 1)$    &	II	&	162   	& 0.037	& $1.2\times10^{2}$     & $8.5\times10^{-3}$      &   $6.07\times10^{2}$   &	$1.10\times10^{-10}$           \\
$(4, -1)$  & $-1/2$  & $(0, 1)$    &	II	&	1828 	& 0.19	& 20  	& $3.9\times10^{-4}$      &    $6.26\times10^{2}$ &	$7.82\times10^{-11}$         \\
$(4, 0)$  & $-1/2$   & $(0, 1)$    &	II	&	165   	& 0.037	& $1.0\times10^{2}$  	& $7.7\times10^{-3}$      &    $5.88\times10^{2}$   &	$1.07\times10^{-10}$           \\
$(4, 1)$  & $-1/2$   & $(0, 1)$    &	II	&	167   	& 0.033	& 75  	& $6.5\times10^{-3}$      &     $6.27\times10^{2}$  &	$1.13\times10^{-10}$          \\
$(4, 2)$  & $-1/2$   & $(0, 1)$    &	II	&	169   	& 0.026	& 47  	& $5.3\times10^{-3}$      &    $7.51\times10^{2}$   &	$1.29\times10^{-10}$           \\
$(4, 3)$  & $-1/2$  & $(0, 1)$     &	II	&	171   	& 0.015	& 21  	& $4.3\times10^{-3}$      &      $1.10\times10^{3}$  &	$1.56\times10^{-10}$           \\
\end{tabular}
\end{ruledtabular}
\end{table*}

For Cs interacting with fermionic isotopes of Yb, resonances can be driven by
any of the three mechanisms discussed in Sec.\ \ref{sec:mechanisms}. This
provides more resonances than for bosonic isotopes, particularly at low field.
The resonances that meet the criteria described above are listed in Table
\ref{resonances_171} for $^{171}$Yb and Table \ref{resonances_173} for
$^{173}$Yb. Each entry in the Tables represents a set of closely spaced
resonances corresponding to different values of $m_{i,\textrm{Yb}}$ (and $M_L$
for mechanism III), as described in Sec.\ \ref{sec:mechanisms}. For each set,
only the widest is given. Full tabulations of the resonances, including all
those in each set and those that are excluded from Tables \ref{resonances_171}
and \ref{resonances_173} by one or more of the criteria, are given in the
Supplemental Material.

The resonances for $^{171}$Yb follow similar patterns to those for $^{173}$Yb,
discussed in Sec.\ \ref{sec:mechanisms}. For $^{171}$Yb, there is a group of
resonances around 74~G caused by bound states with $n=-1$ crossing thresholds
with the same value of $f$. These are all caused by mechanism II. The
corresponding resonances from $n=-2$ states start around 900~G. The remaining
resonances arise from bound states with $f=4$ crossing $f=3$ thresholds, and
arise from mechanisms I and II. The $n=-5$ bound state with $f=4$ lies
approximately 360 MHz below the $f=3$ threshold at zero field; it causes
resonances starting around 150~G. At each threshold, there are resonances of
this type with $\Delta m_f=+1$, 0 and $-1$, at progressively increasing fields,
though not all of them meet the criteria for inclusion in Table
\ref{resonances_171}.

Most of the resonances for fermionic Yb are subject to decay. Tables
\ref{resonances_171} and \ref{resonances_173} include values of the resonant
scattering length $a_\textrm{res}$ and the lifetime $\tau$ that characterize
this decay \footnote{For calculations at finite collision energy with
$L_\textrm{max}>0$, there is always some scattering from $L=0$ into the
degenerate $L=2$ channel, which results in finite $a_\textrm{res}$ even when
there are no truly inelastic open channels. In this case, $a_\textrm{res}$ is
nevertheless infinite at zero energy and the molecular lifetime is also
infinite, even where the Tables list large finite values arising from our
calculations at 100~nK $\times\ k_\textrm{B}$.}. Many of the resonances at
$f=4$ thresholds have $a_\textrm{res}<100\ a_0$ and are likely to be difficult
to observe in loss spectroscopy.

Resonances due to mechanism III are included in Tables \ref{resonances_171} and
\ref{resonances_173}. For $^{171}$Yb, only 3 resonances meet the criteria for
inclusion. For $^{173}$Yb, there are none that meet the criteria, so we have
included the widest undecayed resonance, at 553~G. Resonances due to mechanism
III at excited thresholds are strongly decayed, with $a_\textrm{res} < 10\
a_0$, as exemplified by the resonance at 113~G for Cs (3,2) interacting with
$^{171}$Yb. Such resonances are unlikely to be observable in loss spectroscopy
because $a(B)$ deviates so little from its background value.

There are several resonances in Tables \ref{resonances_171} and
\ref{resonances_173} for Cs ($f=3$, $m_f=3$) interacting with each of
$^{173}$Yb and $^{171}$Yb. These resonances occur at fields where
$\alpha_\textrm{Cs}$ is large, so that experiments in an optical trap are
likely to be hampered by fast intraspecies 3-body losses. However, they would
be good candidates for magnetoassociation in an optical lattice. The strongest
resonances in this category are those at 148~G for $^{171}$Yb and at 619~G and
700~G for $^{173}$Yb.

Cs atoms in magnetically excited states offer additional possibilities.
Promising candidates for observation in loss spectroscopy include those near
$202$~G and $423$~G for $^{171}$Yb and those near $165$~G, $720$~G and 1004~G
for $^{173}$Yb. The resonance near $165$~G for $m_f=2$ has a width similar to
that near $168$~G for $m_f=3$, but $\alpha_\textrm{Cs}$ is much smaller,
corresponding to much slower 3-body loss. The 2-body loss rate $k_2$ is also
very small.

\section{Conclusion} \label{conclusion}

We present a comprehensive theoretical study of magnetically tunable Feshbach
resonances in ultracold collisions between Cs and Yb atoms. We carry out
coupled-channel calculations of the complex scattering length and analyze the
results to obtain resonance positions and widths. For resonances in collisions
of Cs in magnetically excited states, we also extract parameters that
characterize resonance decay and the lifetime of the molecular states
responsible for the resonances.

We use an accurate interaction potential recently determined from
photoassociation spectroscopy \cite{CsYb_02photon}, which gives reliable
scattering lengths for all isotopic combinations of Cs and Yb and gives
accurate predictions for the energies of the molecular states that cause
Feshbach resonances.

The resonances are driven by couplings due to spin-dependent terms in the
Hamiltonian that vary with the internuclear distance. We carry out electronic
structure calculations of the distance-dependence of all the important
spin-dependent interactions, including the scalar hyperfine, tensor hyperfine,
nuclear electric quadrupole, and spin-rotation terms. The resulting couplings
allow us to make quantitative predictions of resonance widths and other
properties.

For bosonic isotopes of Yb, with zero nuclear spin, the resonances are driven
almost entirely by the distance-dependence of the scalar hyperfine interaction
on Cs. The general features of the resulting resonances have been explored in
previous work, but the much improved interaction potential used here allows us
to make specific predictions of the resonance positions and widths for the
first time.

For fermionic isotopes of Yb, with non-zero nuclear spin, there are several
additional terms in the hyperfine Hamiltonian, including significant
anisotropic terms that couple atomic and molecular states with different values
of the partial-wave (or molecular rotation) quantum number $L$. The additional
terms cause additional Feshbach resonances. They also split both the atomic and
molecular states: the atomic states are split into regularly spaced Zeeman
components, but the molecular states are split in more complicated ways,
particularly for $L>0$, and several different spin-dependent terms contribute.
Each Feshbach resonance that would exist in the absence of these terms is split
into a closely spaced set of resonances, spread over 1~G or less.

A particular feature of the fermionic systems is that bound states below one Cs
$f=3$ threshold can cause resonances at another $f=3$ threshold with a
different value of $m_f$. Because these states can be very weakly bound, they
can cause resonances at relatively low field.

We have made a complete set of predictions for all Feshbach resonances below
5000~G for all isotopic combinations. We have identified resonances that are
particularly promising for experimental investigation, both to detect
resonances in an optical trap and to form molecules by magnetoassociation in an
optical lattice.

\begin{acknowledgments}
We are grateful to Florian Schreck for valuable discussions. This work was
supported by the U.K. Engineering and Physical Sciences Research Council
(EPSRC) Grants No.\ EP/I012044/1, EP/N007085/1, EP/P008275/1 and EP/P01058X/1.
JA acknowledges funding by the Spanish Ministry of Science and Innovation Grant
No.\ CTQ2015-65033-P. PS\.Z acknowledges funding from National Science Center
(NCN) Grant no.\ 2017/25/B/ST4/01486.
\end{acknowledgments}

\bibliography{CsYb_res,../../all}

\end{document}